\documentclass[%
 aps,
 preprint,
 amsmath,amssymb,
]{revtex4-2}

\usepackage{graphicx}
\usepackage{dcolumn}
\usepackage{bm}
\usepackage{subfigure}
\usepackage{multirow}
\usepackage{siunitx}
\usepackage{comment}

\begin{document}
\preprint{APS/123-QED}
\title{Determination of the $^{209}$Bi(n,$\gamma$)$^{210g}$Bi cross section using the NICE-Detector}



\author{K. Al-Khasawneh}
\email{kafa.khasawneh@gmail.com}
\author{E.~Borris}
\author{B.~Br\"uckner}
\author{P.~Erbacher}
\author{S.~Fiebiger}
\author{K.~G\"obel}
\author{T.~Heftrich}
\author{T.~Kisselbach}
\author{D.~Kurtulgil}
\author{C.~Langer}
\author{M.~Reich}
\author{R.~Reifarth}
\author{B.~Thomas}
\author{M.~Volknandt}
\author{M.~Weigand}

\affiliation{%
Goethe University Frankfurt, Germany\\
}%


\author{K. Eberhardt}

\affiliation{
 Johannes Gutenberg University Mainz, Germany\\
}%
\author{R. Gernh\"auser}
\affiliation{%
 Technical University Munich, Germany\\
}%


\date{\today}

\begin{abstract}
The capture cross section of $^{209}$Bi(n,$\gamma$)$^{210g}$Bi was measured at different astrophysically energies including thermal capture cross section (25 meV), resonance integral, and the Maxwellian averaged cross section  at a thermal energy of $kT$ = 30 keV. The partial capture cross section ($\sigma_g$) was determined using the activation technique and by measuring the $^{210}$Po activity.  The newly developed and tested NICE detector setup was used to  measure the $\alpha$-activity of the $^{210}$Po.  Using this setup the thermal and resonance integral cross sections were determined to be $16.2\;\pm\;0.97$~mb and $89.81\;\pm\;8.0$~mb, respectively. And the Maxwellian  average cross section was measured to be $2.01\;\pm\;0.38$~mb.
\end{abstract}
\maketitle

%
%
%
%
%


\section{Introduction}
 Accurate data of the $^{209}$Bi(n,$\gamma$)$^{210}$Bi cross section are necessary to explain the elemental abundances near the s-process termination point. Furthermore, capture cross section values have become an essential matter after considering Pb-Bi as a coolant material in the fast reactors and as a target material in the Accelerator-Driven Systems (ADS). 

\begin{figure}[htb]
\centering 
\includegraphics[scale=0.4]{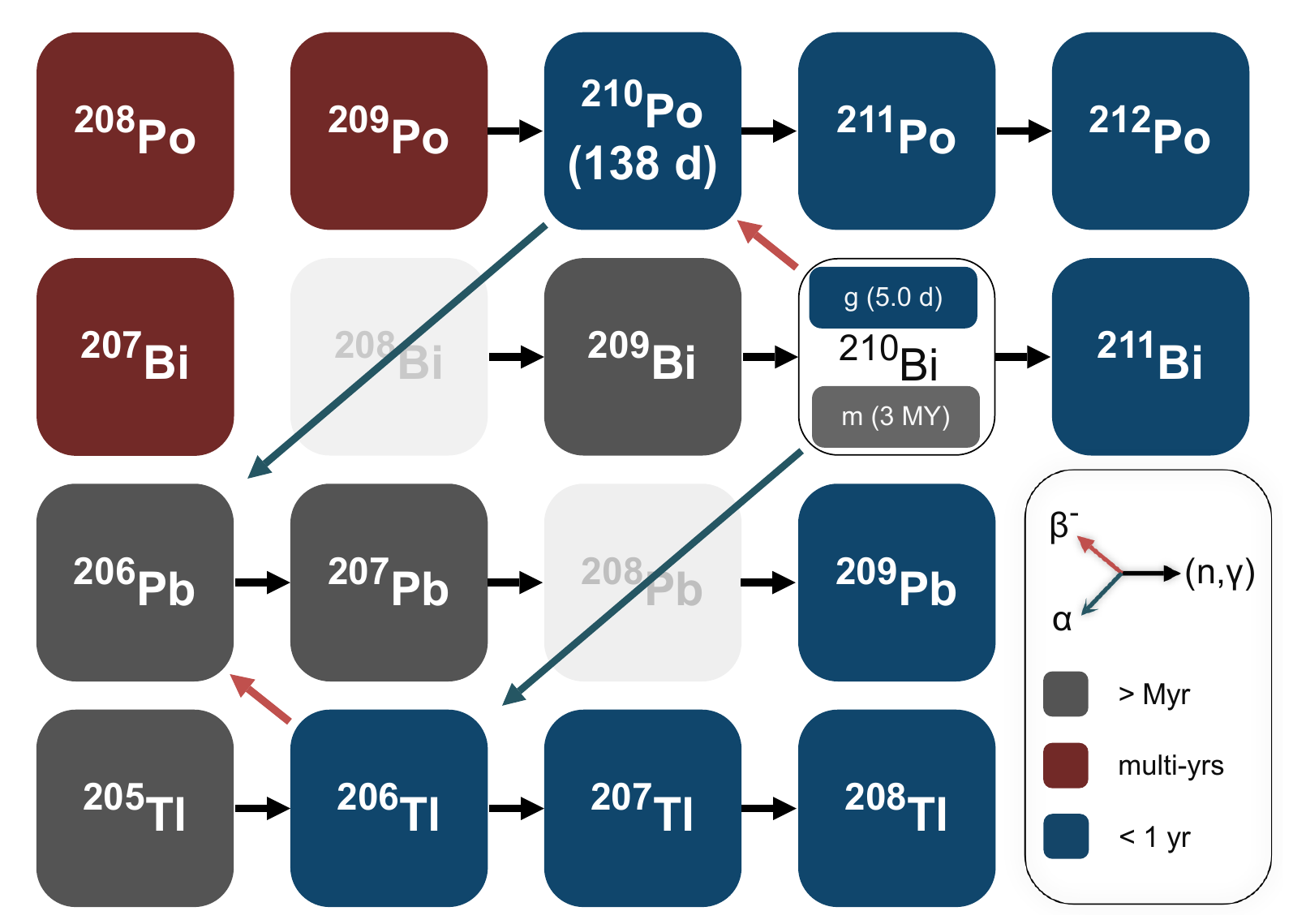}
\caption[sprocess.]{s-Process near the termination point.}
\label{fig:sBi}
\end{figure}

According to the s-process scenario, $^{209}$Bi is the heaviest stable isotope (or long-lived, t$_{1/2}$ $\simeq$ 10$^{19}$ Y \cite{Marcillac}). Since the unstable nuclei between  $^{209}$Bi and the meta-stable Th/U isotopes can not be overcome during s-process conditions, $^{209}$Bi resembles the end point of the s-process. Figure \ref{fig:sBi} depicts the s-process near Bi, the neutron capture on $^{209}$Bi leads to  the production of  $^{210}$Bi  in either its ground state $^{210g}$Bi or in the long-lived state $^{209m}$Bi ($E$ = 271.3 keV). All nuclei produced in their ground state undergo $\beta$-decay ($t_{1/2}$ = 5.03 days)  to feed the $\alpha$-unstable $^{210}$Po isotope, which terminates the s-process chain and recycles its flow back to $^{206}$Pb by emission of 5.3~MeV $\alpha$-particles ($t_{1/2} \;\simeq\; 138$ days). $^{210}$Po with a relatively long half-life can capture another neutron and contribute to the  production of $^{207}$Pb. On the other hand, the long-lived isomer state ($t_{1/2}$ = 3.04 $\times$ 10$^{6}$ Y) can also capture a neutron and lead to the production of $^{211}$Bi, which undergoes $\alpha$-decay into $^{207}$Tl.  

A number of experimental data of $^{209}$Bi(n,$\gamma$)$^{210g}$Bi capture cross section at different neutron energies were reported in previous studies, including thermal neutrons \cite{seren, Colmer, Takiue, Letourneau}, neutrons in the resonance region \cite{Macklin, Domingo, Mutti, Beer}, and neutrons with quasi-Maxwellian distribution at $kT$ = 30 keV \cite{Ratzel,Bisterzo,Shor}. With  an overview of this data and when compared to the evaluated values,  one can see a considerable disagreement. Due to this discrepancy and for comparison purposes, new data are necessary. 
 
In this study, The cross section was determined using the activation technique and by measuring the $^{210}$Po activity. For this purpose, a new detector setup (NICE-Neutron Induced Charge particle Emission) was developed and tested  to be used for measuring the $\alpha$-activity. Before going through the cross section calculation, the NICE detector design and performance are presented first. 
\section{The NICE Detector}\label{NICE}
The NICE detector was designed to be used in experiments of neutron-induced reactions with a charged particle in the exit channel, including activation and Time-of-Flight  technique (ToF). In these setups, the charged particles measurement would be performed in an environment full of background radiation, including electrons, $\gamma$~radiation. Therefore, the NICE detector should have high sensitivity to measure charged particles and low sensitivity for background radiation. In addition, a fast-timing detector is also required, so it can recover very fast from the huge background count rate. 


\subsection{The NICE detector design}
\begin{figure*}[!htb]
\centering
\subfigure[]{\includegraphics[scale=0.6]{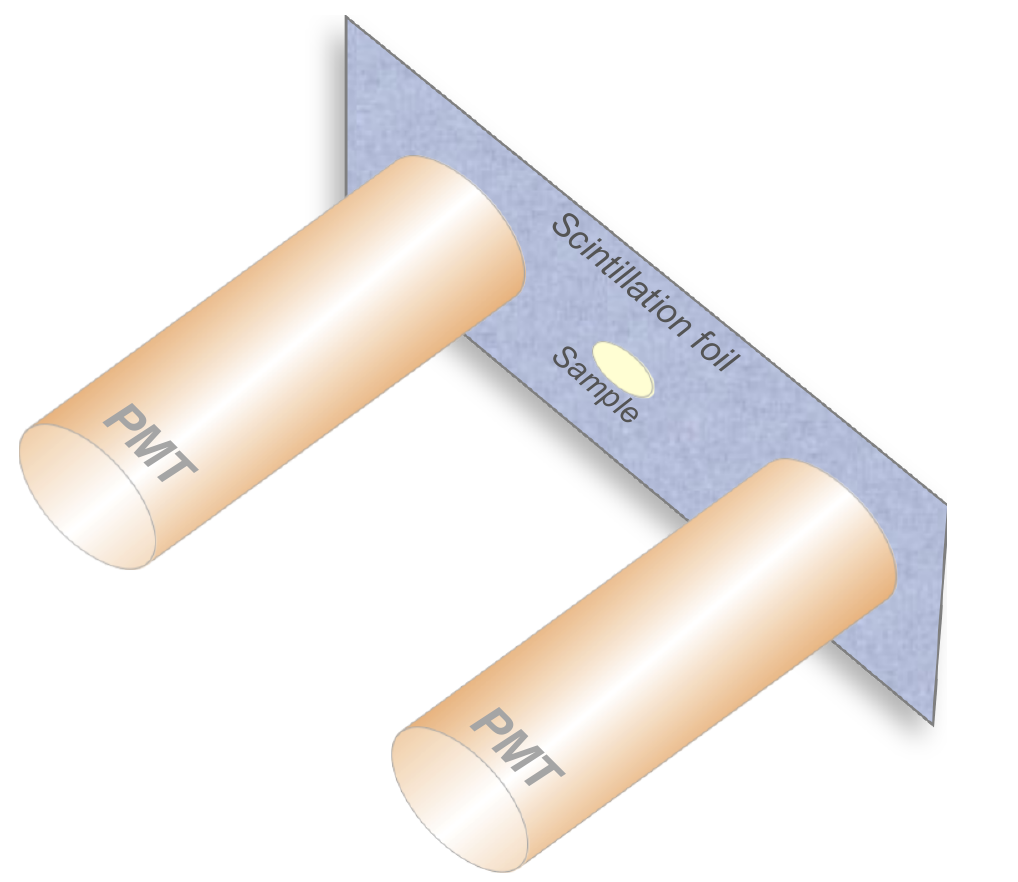}}
\subfigure[]{\includegraphics[scale=0.05]{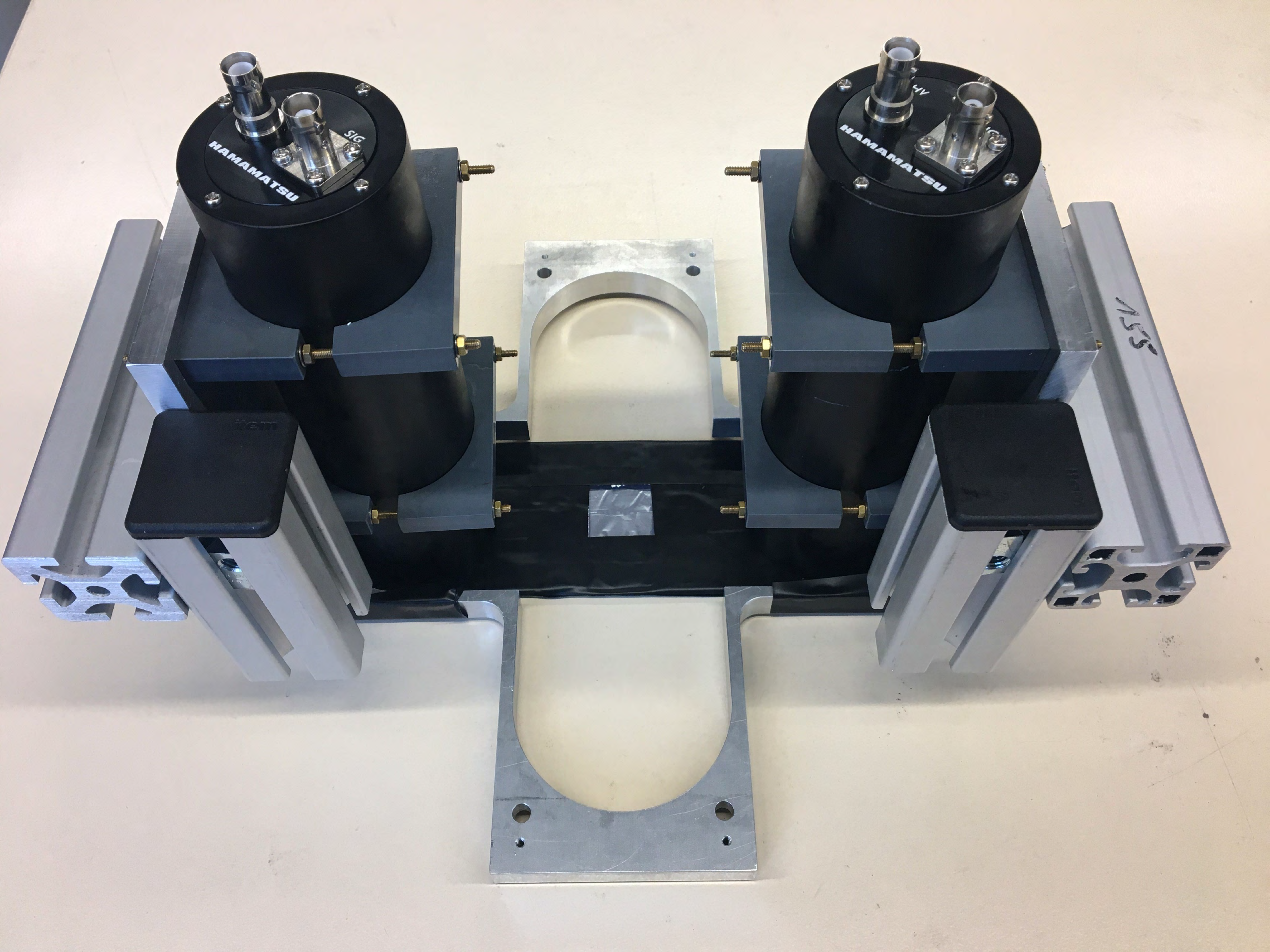}}
\caption{(a) Schematic design represents the NICE detector components and the expected neutron beam direction. The dimensions of the scintillator foil are 26~$\times$~7~cm$^2$ and the PMTs have a diameter of 6.0~cm. (b) The NICE detector and the PMTs supporting structure.}
\label{NICE-detector}
\end{figure*}

The NICE detector design is composed of a thin layer of plastic scintillator, coupled to two Photomultiplier Tubes (PMT) at one face of the scintillator foil and connected to readout electronics (Figure \ref{NICE-detector}).
According to this flexible design, the scintillator thickness can be varied for each experiment. Ideally, the chosen thickness should not exceed the range of the charged particles in the scintillator material, thereby, one can obtain a high detection efficiency for the charged particles and reduce the contribution of the background radiation into negligible levels. In this work, the scintillator in use is a polyvinyltoluene (PVT) based material with the manufacturer's product code BC-408 \cite{SG}, which has 26~$\times$~7~cm$^2$ surface area and $0.1$ mm thickness. The PMTs are type H2431-50  from Hamamatsu \cite{PMT} with a diameter of 6.0~cm. The NICE detector schematic and final design are shown in Figure \ref{NICE-detector}. 

\subsection{Alpha detection efficiency of the NICE detector }\label{NICE-DetectorCharacteristic study}
The detection efficiency of the NICE detector was determined using an Americium radioactive source $^{241}$Am ($E_{\alpha}$= 5.5 MeV, $I_{\alpha}$ = 100$\%$). A point source of activity 1.17~kBq was placed at the center and in close contact with the scintillation surface to avoid any measurable energy loss in the air.

\begin{figure*}[!htb]
\subfigure[]{\includegraphics[scale=0.4]{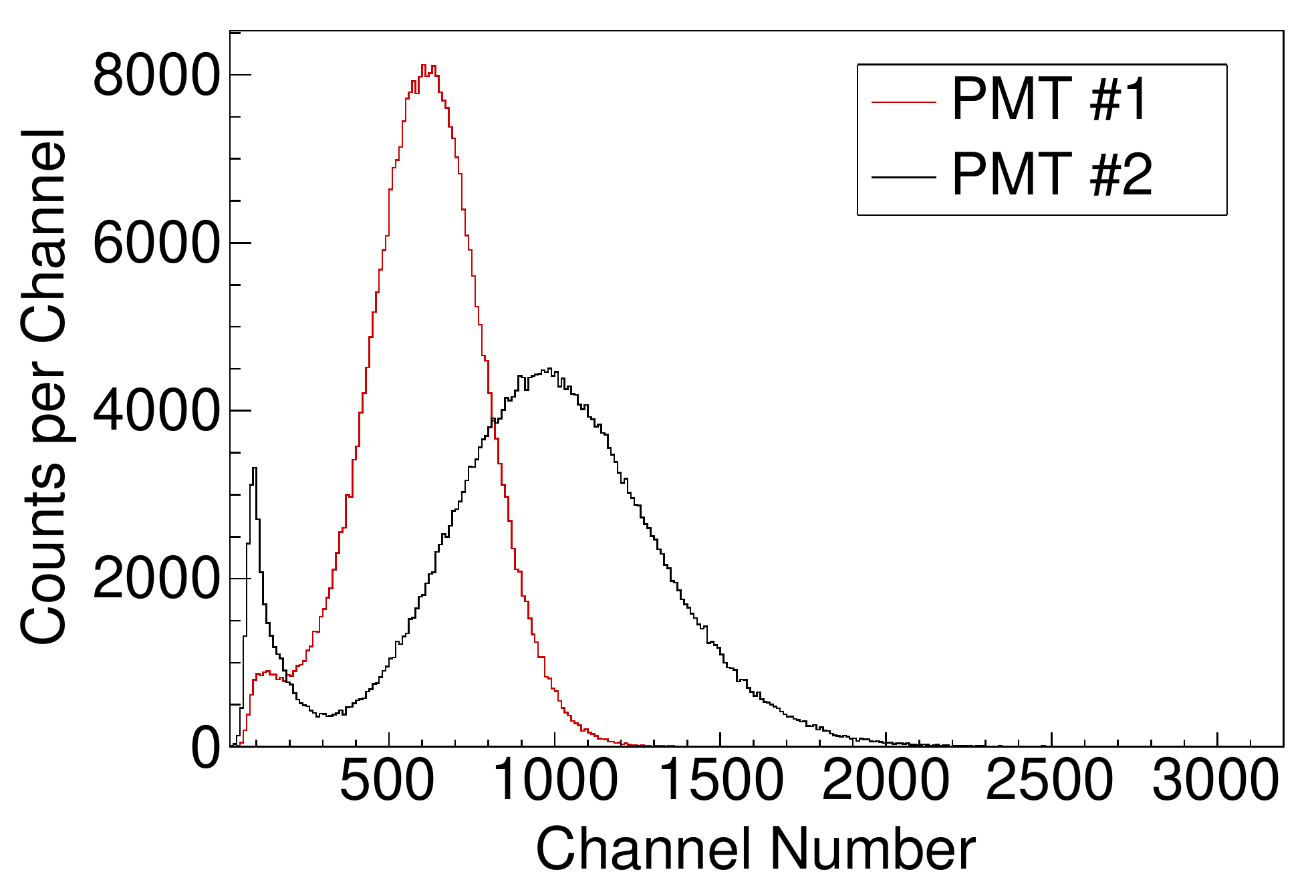}}
\subfigure[]{\includegraphics[scale=0.4]{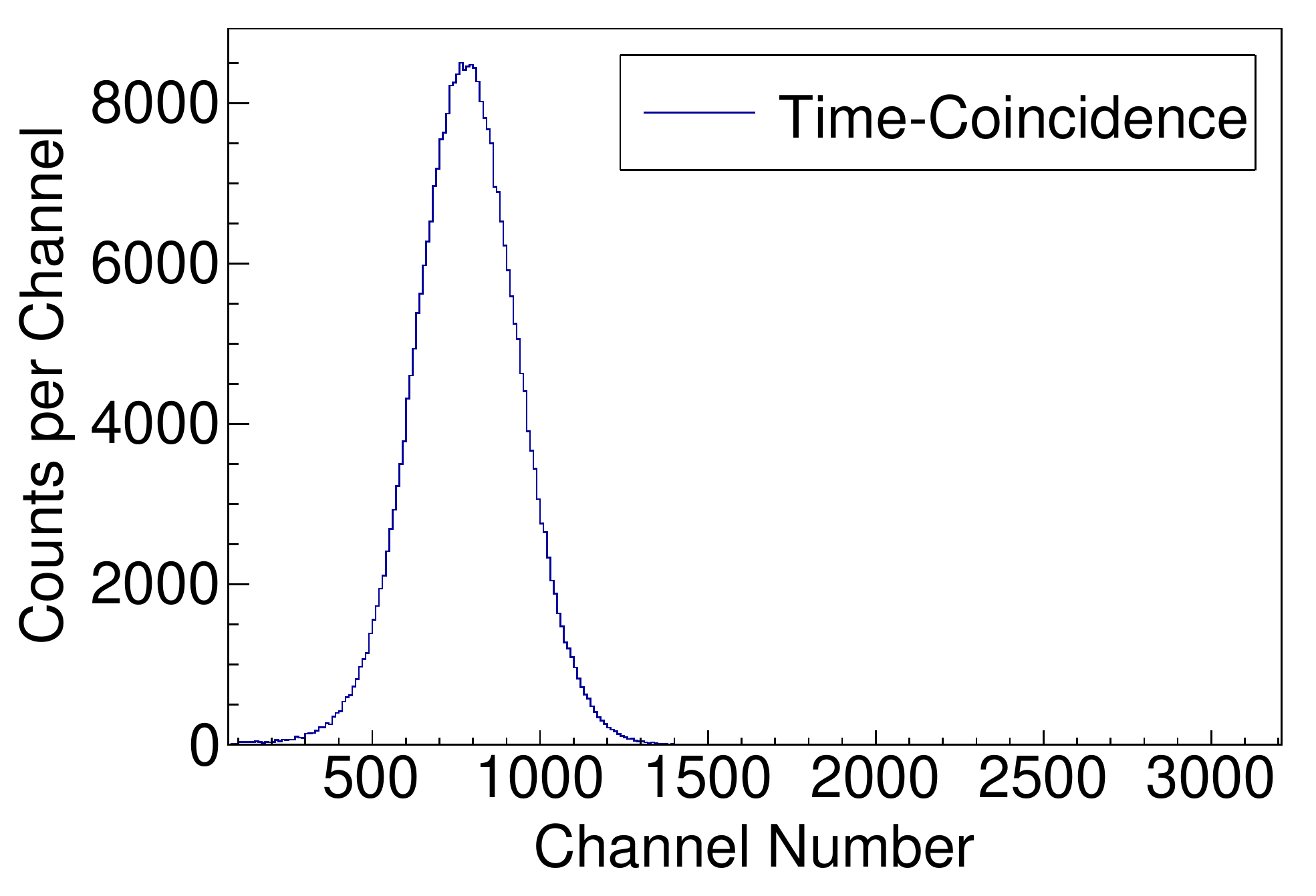}}
\caption{(a) Pulse height distribution of $^{241}$Am source obtained via PMT1 and PMT2. The different peak position is due to the different total gain. (b) Pulse height distribution after applying  time-coincidence condition.}
\label{PHD}
\end{figure*}

Figure \ref{PHD} shows the pulse height distribution of both PMTs as a function of channel number (relative light output). Due to the considerable distance between the interaction position and the PMTs centers (= 6.5 cm), light intensity measured by each PMT is relatively low; accordingly, the signal level is very low and overlaps with the electronic noise signals level (Figure \ref{PHD} (a)). To distinguish between real scintillation pulse signals and electronic noise signals, the time coincidence technique using both PMTs was adopted with a 50~ns  coincidence window. The real scintillation pulse from the specific physical event (e.g $\alpha$, $\gamma$, e-) is detected by both PMTs nearly at the same time, while noise and background signals are mostly not correlated. The feasibility of this technique is illustrated in Figure \ref{PHD} (b), which shows the pulse height distribution for the $\alpha$-peak after applying the time coincidence technique, where the noise and background levels are significantly reduced. 
The total detection efficiency was calculated from the total number of counts under the $\alpha$-peak and was found to be $\varepsilon_{\alpha=5.5MeV}$~=(46.31 $\pm$ 0.47)$\%$.
\subsection{Gamma detection efficiency of the NICE detector }\label{NICE-DetectorCharacteristic study_gamma}
The NICE detector setup  and the detection technique were investigated using gamma point-like sources, the experimentally measured spectra were compared to those obtained using the Geant4-simulation model. 
In this work, three standard point-like sources: $^{137}$Cs,  $^{54}$Mn, and $^{133}$Ba were used.
Each source was  placed in contact with the scintillator and counted for an adequate time to obtain  sufficient statistics for comparison with the simulated spectra. An independent background measurement was performed for one  day as well. 

\begin{figure}[t!]
\centering 
\subfigure[]{\includegraphics[scale=0.5]{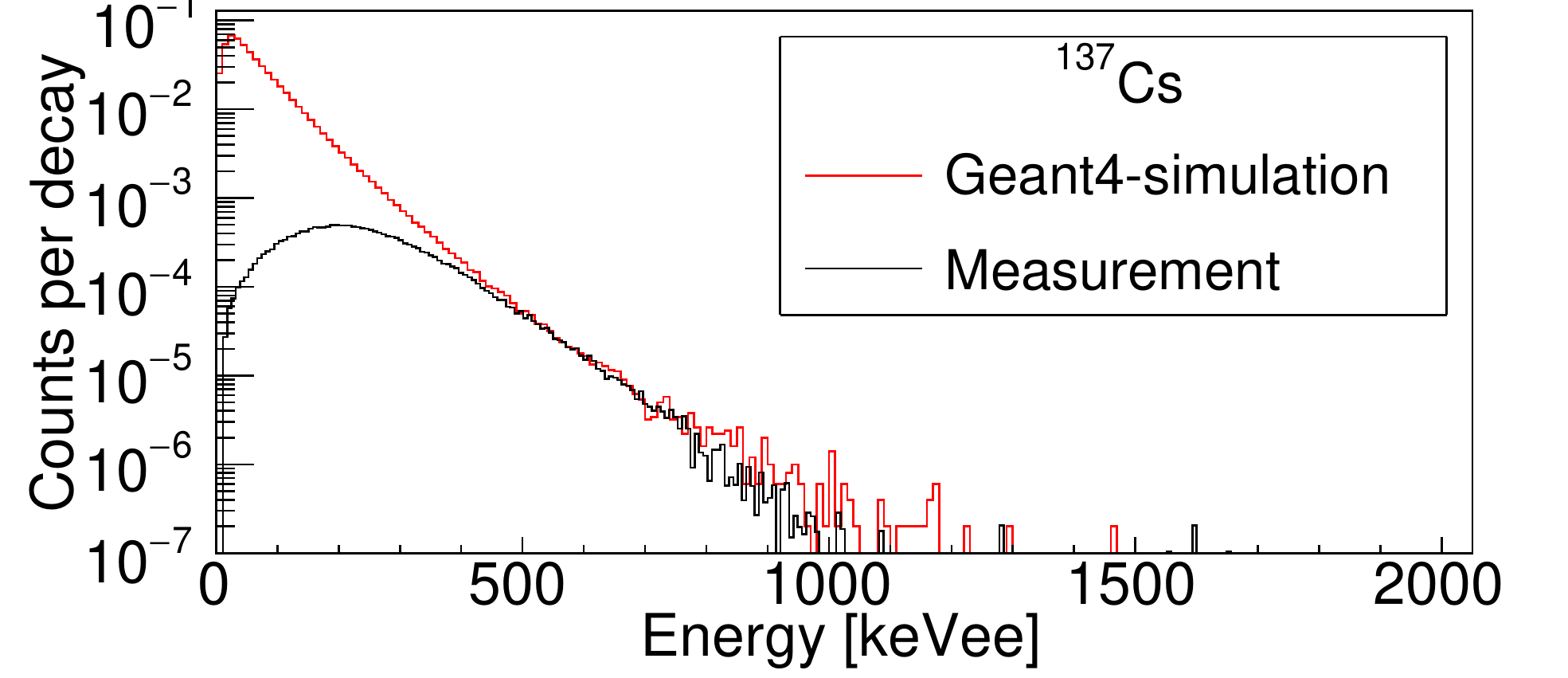}}\\
\subfigure[]{\includegraphics[scale=0.5]{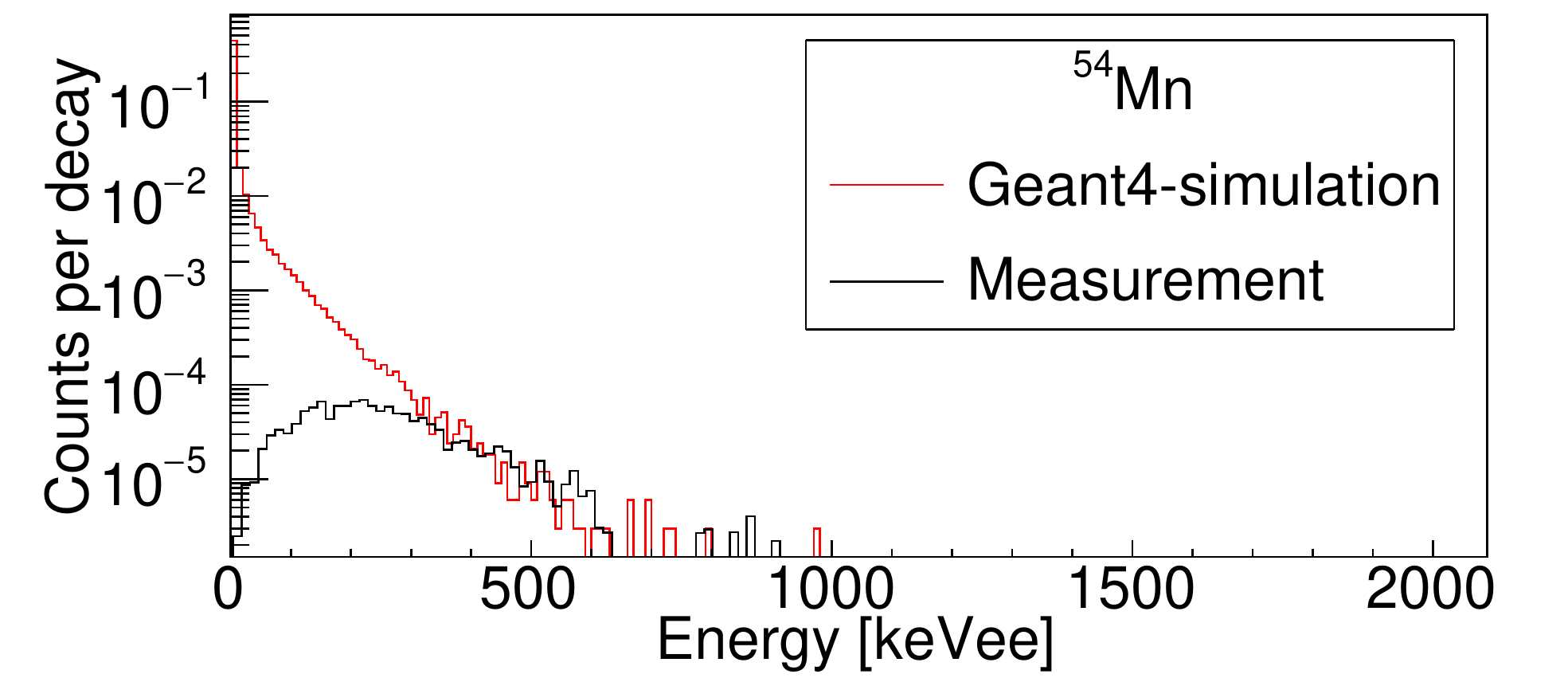}}\\
\subfigure[]{\includegraphics[scale=0.5]{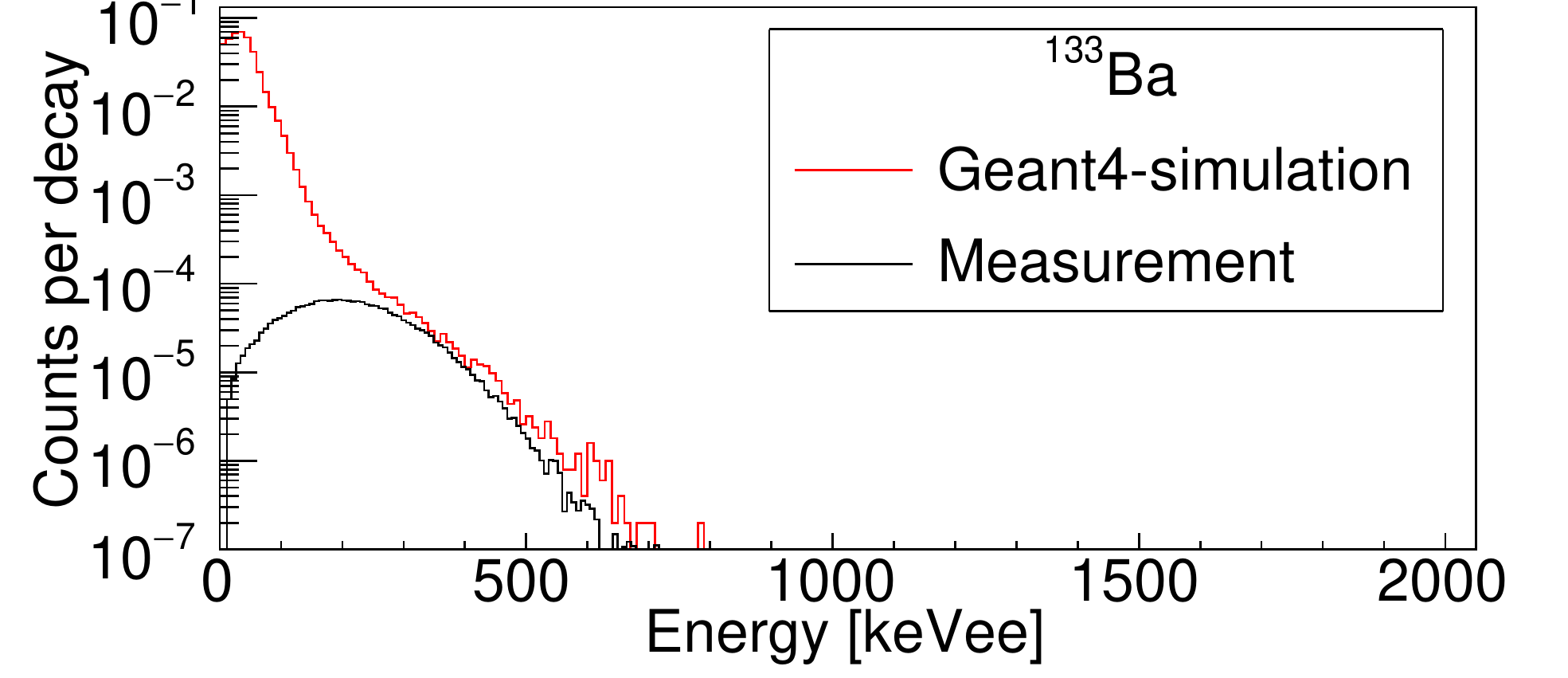}}\\
\caption[The simulated gamma spectrum compared to the calibrated experimental spectrum. (a) $^{137}$Cs, (b) $^{54}$Mn, and (c) $^{133}$Ba.]{The folded simulated gamma spectrum compared to the calibrated experimental spectrum. (a) $^{137}$Cs, (b) $^{54}$Mn, and (c) $^{133}$Ba. The energy is given in terms of keV electron equivalent.}
\label{fig:comp}
\end{figure}

The obtained pulse hight distribution for each source was  normalized per decay after subtracting the background contribution and are illustrated in Figure \ref{fig:comp}. A good agreement between simulated and measured energy spectra for all  sources can be seen. A continuum spectrum is the only feature of the distribution, and one can not identify neither the full peak nor a Compton edge for any of the gamma lines. This is due to both the low detection efficiency for gamma radiation and the relatively poor energy resolution of the NICE detector. The total detection eefficiencies for the three sources were measured and found to be lower than $1\%$.
\\
\\
\\
\section{The $^{209}$B\MakeLowercase{i}(\MakeLowercase{n},\MakeLowercase{$\gamma$})$^{210}$B\MakeLowercase{i} Cross section}\label{Bi Cross section}
\subsection{The Bi-samples}\label{Bi-samples}
In this work, three Bi-samples were used (Bi-I, Bi-II, and Bi-III), each composed of a thin layer of high purity $^{209}$Bi (Purity = 99.97$\%$) sputtered on  polystyrene backing of 0.1 mm thickness  \cite{Bi_goodfellow}. To determine the thermal and resonance integral cross sections the samples Bi-I and Bi-II with a surface area of 1.56 cm$^{2}$ were used and activated using the TRIGA research reactor at Mainz University. The sample  Bi-III has a surface area of 6.25 cm$^{2}$  (2.5 $\times$ 2.5 cm$^2$) and was used to measure the partial capture cross section at thermal energy of $kT$ = 30 keV and was activated using the Van de Graaff accelerator at Frankfurt University. All  sample thicknesses  were determined experimentally based on an X-ray absorption measurement. Table \ref{tab:Bi-Sample Thickness} lists the basic characteristic of each sample and the initial number of atoms. 
\begin{table}[]
 	\begin{center}
   	\caption{Surface area, thicknesses  and the total number of atoms ($N_0$) for the Bi-samples used during the activations}
   	 \label{tab:Bi-Sample Thickness}
  	  \begin{tabular}{ c c  ccc cc} 
   	 \hline
   	\hline				 
\multirow{2}{*}{Sample}\quad& \quad Surface area\quad& \quad Thickness\quad&  \quad$N_{0}$ \quad\\
\quad&\quad [cm$^2$]\quad&\quad$[\mu m$] \quad&  \quad [$10^{19}$ atoms] \quad\quad\\
   	\hline
	 && &\\

Bi-I\quad& \quad1.56\quad& \quad5.35 $\pm$ 0.24\quad&  \quad 2.35 $\pm$ 0.11\quad\\

Bi-II\quad&  \quad1.56\quad& \quad 5.42 $\pm$ 0.35\quad&  \quad 2.37 $\pm$ 0.15\quad\\

Bi-III\quad&  \quad 6.25\quad& \quad5.32 $\pm$ 0.31\quad&  \quad  9.37 $\pm$ 0.54\quad\\
 	&&&\\
     \hline
    \hline
    \end{tabular}
  \end{center}
\end{table}

\subsection{Thermal cross section and resonance integral}

\subsubsection{Activation}
\hfill\\
\\
The activation process took place at the research reactor TRIGA$-$Mark II at Johannes Gutenberg$-$University Mainz, Germany. In this work cross sections were determined using the Cd ratio difference method. Therefore, two activations were performed, once without a Cd filter where the sample was exposed to the full neutron flux spectrum, and once with a Cd filter where the thermal component of the neutron flux will be absorbed by the Cd filter, leaving only the epithermal component available for the sample.   Wires of Al-0.1$\%$Au alloy, and natural foils of $^{45}$Sc (thickness = 0.1 mm) were used to determine the thermal and epithermal component of the neutron flux. Both gold and scandium are considered as excellent flux monitors, because they can be produced with high purity, have relatively high capture cross section and  convenient half-lives. The corresponding decay data for both reactions are listed in Table~\ref{tab:tabledata}.
\begin{table}[h]\small
 
  \begin{center}
    \caption{Nuclear data for $^{197}$Au and $^{45}$Sc used in this work. Half-lives and emission probabilities  are taken from the Brookhaven NNDC Database \cite{NNDC}, while the neutron capture cross section values are taken from ENDF \cite{ENDF}. }
    \label{tab:tabledata}
    \begin{tabular}{ ccccccc} 
    \hline
    \hline

   \multirow{2}{*}{Reaction }\quad	&\quad $t_{1/2}$\quad		&	\quad $E_{\gamma}$\quad	&\quad $I_{\gamma}$\quad	&\quad$\sigma_{0}$\quad		&\quad $I_{0}$\quad	 \\
        							&\quad [d]\quad		& 	\quad [keV] \quad 			&\quad [$\%$] \quad				 & \quad [barn]\quad			&\quad[barn]\quad			 \\
   \hline
        	&&&\\

   $^{197}$Au (n,$\gamma$)$^{198}$Au\quad 	 &\quad2.69\quad	&\quad411.8\quad	&\quad95.62\quad 	&\quad98.7\quad 	&\quad1571\quad   \\
       	&&&\\

   $^{45}$Sc (n,$\gamma$)$^{46}$Sc\quad	&\quad83.79 \quad	 &\quad889.3\quad	&\quad99.99\quad	&\quad27.2\quad 	&\quad12.06\quad   \\
 	&&&\\

    \hline
    \hline
    \end{tabular}
  \end{center}
\end{table}

\begin{table*}[t]
 	\begin{center}
   	\caption{Masses and the number of atoms ($N_{0}$) for the neutron flux monitors.}
   	 \label{tab:samples}
	 \begin{ruledtabular}
  	  \begin{tabular}  { c c c  | c c c} 
   	
   	 \multicolumn{3}{c|}{{Activation without Cd}}&\multicolumn{3}{c}{{Activation with Cd}}\\
	\hline
   \multirow{2}{*}{ {Monitor}}\quad	&\quad  {Mass} \quad&\quad $ {N_{0}}$ \quad	&\multirow{2}{*}{ {Monitor}} \quad		&\quad  {Mass} \quad	&$ \quad  {N_{0}}$ \quad  \\
    						&\quad  {[mg]} \quad&\quad  {[$ 10^{19}$ atoms]}\quad		&			& 	\quad  {[mg]} \quad		& \quad  {[$10^{19}$ atoms]} \quad  \\
    	  \hline

 	&& &&&\\
	 {Au-1}\quad&\quad {1.1$\times10^{-2}$}\quad&\quad {$(3.33\;\pm\;0.03)$$\times10^{-3}$}\quad& {Au-3}\quad&\quad {1.0$\times10^{-2}$}\quad&\quad {$(3.06\;\pm\;0.03)$$\times10^{-3}$}\quad   \\
	 {Au-2}\quad&\quad {1.1$\times10^{-2}$}\quad&\quad {$(3.33\;\pm\;0.03)$$\times10^{-3}$}\quad& {Au-4}\quad&\quad {1.1$\times10^{-2}$}\quad&\quad {$(3.30\;\pm\;0.03)$$\times10^{-3}$}\quad \\

 	&& &&&\\
	 {Sc-1}\quad&\quad {3.20}\quad&\quad {$4.29\;\pm\;0.07$}\quad& {Sc-3}\quad&\quad {2.7}\quad&\quad {$3.62\;\pm\;0.07$}\quad  \\
	 {Sc-2}\quad&\quad {3.60}\quad&\quad {$4.82\;\pm\;0.07$}\quad& {Sc-4}\quad&\quad {4.2}\quad&\quad {$5.63\;\pm\;0.07$}\quad  \\

    \end{tabular}
    \end{ruledtabular}

  \end{center}
\end{table*}

To take into account the neutron flux gradient, the sample was positioned in-between two monitor sets. Detailed parameters of the  monitors for each  activation are given in Table \ref{tab:samples}.
The samples and flux monitors were sealed with plastic pockets to protect them from any external contamination, then packed into high purity polyethylene vials. Both activations with and without the Cd filter were performed for 20 min.  

The total number of activated nuclei at the end of the activation interval course was expressed using the H$\o$gdahl convention \cite{Hogdahl} as
\begin{equation}
N_{\SI{}{activation}}= N_{0} (\sigma_{0} \phi_{th}  + I_{0} \phi_{epi}),
\label{nactivation}
\end{equation}
where $N_{0}$ is the initial number of sample atoms, $\phi_{th}$ and $\phi_{epi}$ are the time integrated thermal and epithermal neutron flux (n/cm$^{2}$), respectively. $\sigma_{0}$ and $I_{0}$ are thermal cross section and resonance integral, respectively.

\subsubsection{Calculations of $\phi_{th}$ and $\phi_{epi}$}
\hfill\\
\\
Induced activities for flux monitors were measured using HPGe-detector. The total number of activated nuclei ($N_{\SI{}{activation}}$) is described as \cite{activation}  
 \begin{equation}
N_{\SI{}{activation}}= \frac{C_{\gamma}}{\varepsilon_{\gamma} I_{\gamma} k} \quad \frac{\lambda_{\gamma} t_{a}}{1- e^{-\lambda t_{a}}} \quad\frac{1}{e^{-\lambda_{\gamma} t_{w}}} \quad\frac{1}{1- e^{-\lambda_{\gamma} t_{m}}},
\end{equation}

where  $C_\gamma$ is the number of gamma counts in a particular gamma-line,  $\varepsilon_{\gamma}$ is the detector efficiency,  $\lambda_{\gamma}$  is the decay constant of the respective product nucleus, $I_{\gamma}$ is the relative emission probability  for a particular gamma energy,  $k$ is the dead-time correction factor,  and determined from the ratio of real-to-live time, $t_{a}$, $t_{w}$, $t_{m}$ are the activation, waiting  and  measurement times, respectively.  

With the assumption that both flux monitors were exposed to the same neutron flux during the activation time course, $\phi_{th}$ and $\phi_{epi}$ can be calculated as follows
\begin{eqnarray}
\phi_{th}&= \frac{[\frac{^{^{198}\SI{}{Au}}{N}}{^{^{197}\SI{}{Au}}{N}}]I_{0}^{\SI{}{Sc}}-[\frac{^{^{47}\SI{}{Sc}}{N}}{^{^{46}\SI{}{Sc}}{N}}]I_{0}^{\SI{}{Au}}}{\sigma_{0}^{\SI{}{Au}}I_{0}^{\SI{}{Sc}}-\sigma_{0}^{\SI{}{Sc}}I_{0}^{\SI{}{Au}}},\\
\phi_{epi}&= \frac{[\frac{^{^{198}\SI{}{Au}}{N}}{^{^{197}\SI{}{Au}}{N}}]\sigma_{0}^{\SI{}{Sc}}-[\frac{^{^{47}\SI{}{Sc}}{N}}{^{^{46}\SI{}{Sc}}{N}}]\sigma_{0}^{\SI{}{Au}}} {\sigma_{0}^{\SI{}{Sc}}I_{0}^{\SI{}{Au}}-\sigma_{0}^{\SI{}{Au}}I_{0}^{\SI{}{Sc}}}.
\end{eqnarray}
Table \ref{tab:table1} gives the total number of activated nuclei, time integrated flux values, and the corresponding average flux values seen by  the Bi-samples for both activations. It also provides the statistical uncertainties determined from the total number of counts under the gamma-peak ($\Delta_{stat} = \sqrt{C_\gamma}$), while systematic uncertainties originated from the $\gamma$-efficiencies,  decay intensities, half-lives and cross section values.
\begin{table*}[t]

  \begin{center}
    \caption{The total number of activated nuclei ($N_{\SI{}{activation}}$) determined by each flux monitor for both activation. And the time integrated neutron fluxes measured by each flux monitor set, and the time integrated neutron flux seen by the Bi-sample for both activations.}
    \label{tab:table1}
     \begin{ruledtabular}
    \begin{tabular}{ c cc c c } 
   
     \multirow{2}{*} {Monitor}		&	$N_{\SI{}{activation}}\pm\; \Delta_{stat}\pm\;\Delta_{sys}$	 	&\quad\quad {$\phi_{th}\pm\; \Delta_{stat}\pm\;\Delta_{sys}$} \quad\quad		 &\quad\quad	 {$\phi_{epi}\pm\; \Delta_{stat}\pm\;\Delta_{sys}$ }	\quad\quad		     \\
       
				&	[10$^9$ atoms]		&\quad\quad  {[10$^{13}$ n/cm$^2$]}\quad\quad					  	&\quad\quad	 {[10$^{13}$ n/cm$^2$] }\quad	\quad	       \\
	\hline
	
	\multicolumn{2}{l}{\textbf{Activation without Cd}}\\

    Au-1	& 		3.52 $\pm$ 0.03 $\pm$ 0.06			&\quad\quad\multirow{2}{*}{59.3 $\pm$ 0.38 $\pm$ 1.24}\quad \quad		&\quad\quad\multirow{2}{*}{3.00 $\pm$ 0.07 $\pm$ 0.20}\quad\quad  \\
    Sc-1  	& 		706.54 $\pm$ 4.29 $\pm$ 8.64												&						&\\
 &&&\\
   Au-2	&		3.15 $\pm$ 0.03 $\pm$ 0.05			&\quad\quad\multirow{2}{*}{54.9 $\pm$ 0.36 $\pm$ 1.07}\quad\quad		&\quad\quad\multirow{2}{*}{2.56 $\pm$ 0.06 $\pm$ 0.14} \quad\quad \\
   Sc-2	&		734.91 $\pm$ 4.53 $\pm$ 8.98				& 							&\\
&&&\\
   \multicolumn{2}{l}{\textbf {Average Flux}}   					&\quad\quad\textbf {57.1 $\pm$ 0.26 $\pm$ 0.82}\quad \quad								&\quad\quad\textbf {2.78 $\pm$ 0.05 $\pm$ 0.10} \quad\quad\\
        &&&\\
  \hline
  
  \multicolumn{4}{l}{\textbf{Activation with Cd}}\\

   Au-3	& 		1.39 $\pm$ 0.02 $\pm$ 0.04		&		\multirow{2}{*}{0.021 $\pm$ 0.021 $\pm$ 0.049} &		\multirow{2}{*}{2.89 $\pm$ 0.03 $\pm$ 0.08} \\
   Sc-3  	 & 		12.80 $\pm$ 0.16 $\pm$ 0.19		&															&\\
 &&&\\
   Au-4 	 &		1.40 $\pm$ 0.02 $\pm$ 0.04	&		\multirow{2}{*}{0.030 $\pm$ 0.02 $\pm$ 0.04} 	&\multirow{2}{*}{2.69 $\pm$ 0.03 $\pm$ 0.08}  \\
   Sc-4	 &		18.73  $\pm$ 0.19$\pm$ 0.28	&										&\\
 &&&\\
   \multicolumn{2}{l}{\textbf {Average Flux}}   		&			\textbf {0.025 $\pm$ 0.014 $\pm$ 0.032} 									&\textbf {2.79 $\pm$ 0.02 $\pm$ 0.06} \\
        &&&\\
  
    \end{tabular}
      \end{ruledtabular}

  \end{center}
\end{table*}

\subsubsection{Calculations of $\sigma_{0}$ and $I_{0}$}
\begin{figure}[!htb]
 \centering
\subfigure[]{\includegraphics[scale=0.5]{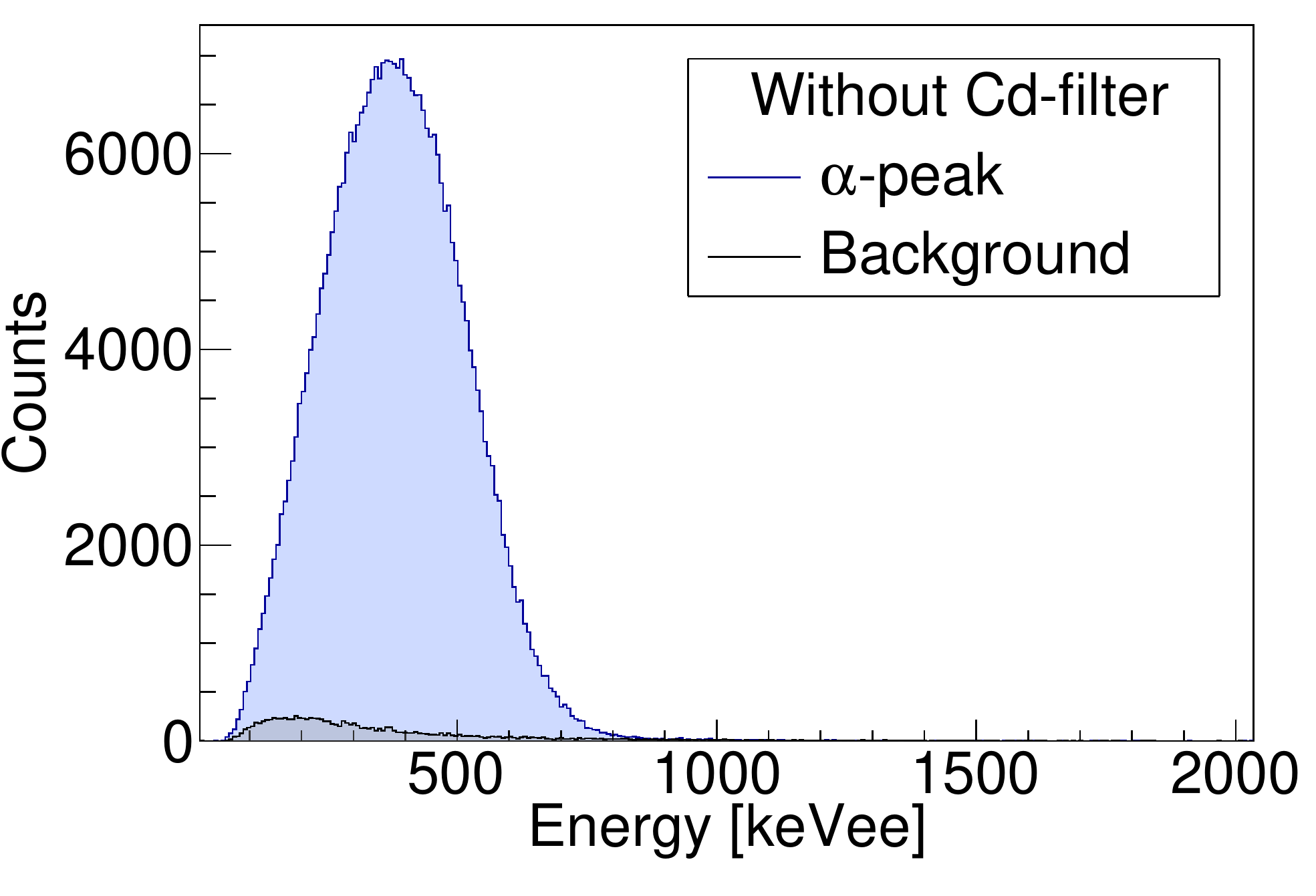}}
\subfigure[]{\includegraphics[scale=0.5]{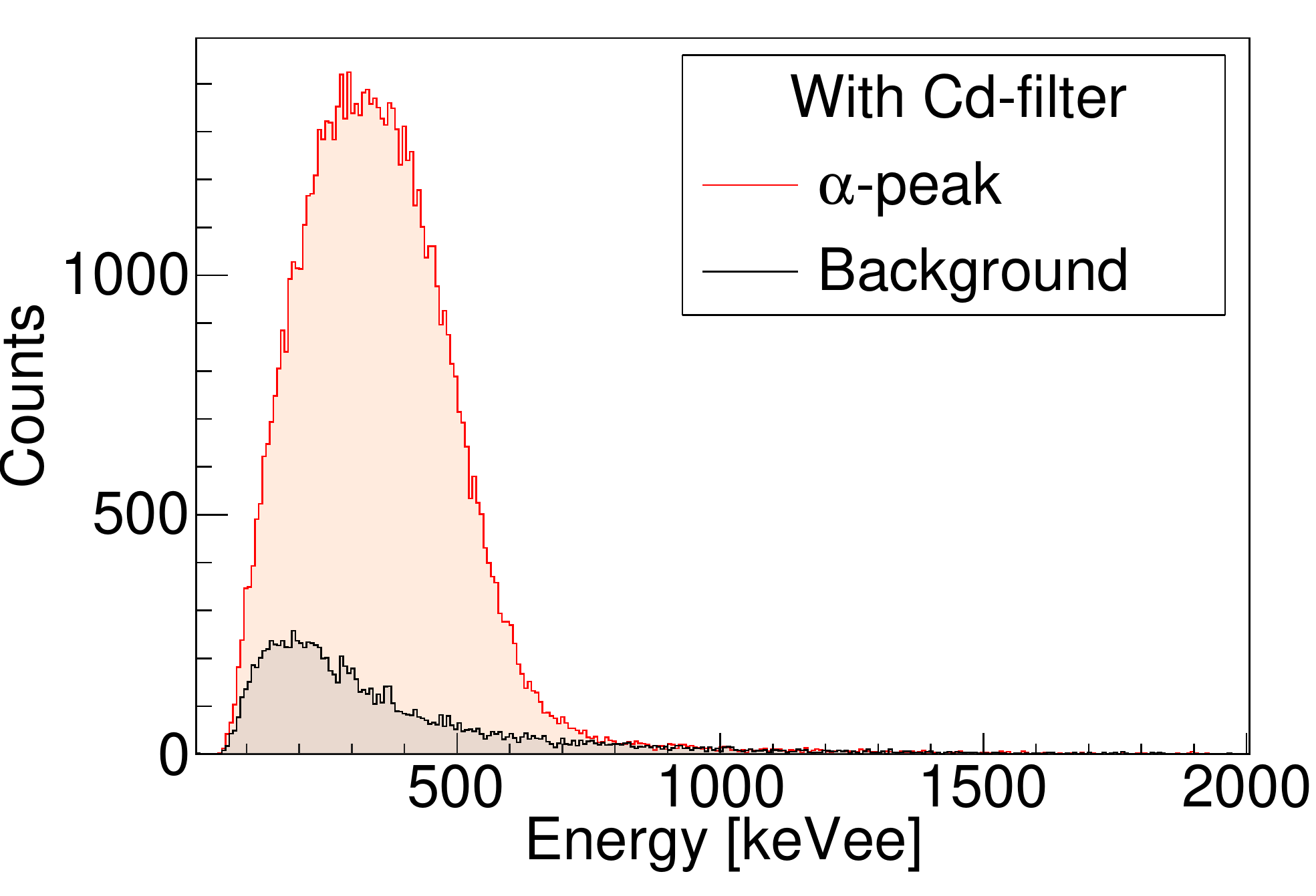}}
\caption{ The $\alpha$-peak obtained using the NICE-detector for the activated bismuth samples, both spectra obtained on day 77 after the  activation and for 24 hours.  (a) Activation without Cd-filter, (b) Activation with Cd-filter.}
\label{Bi_Peak}
\end{figure}

\hfill\\
\\
The induced activities of Bi-samples were measured using the NICE detector. Since the $^{210}$Po half-life is long relative to that of $^{210}$Bi, waiting time in the range of $\simeq$ 77 days was sufficient to reduce the $^{210}$Bi activity ($\beta^-$ activity) into negligible levels, and at the same time reach a measurable level of the $^{210}$Po activity. In order to investigate the $\alpha$-peak decay behaviour, 35 consecutive measurements were performed each for 24 hours.  An example of  a typical $\alpha$-peak obtained from one measurement is given in Figure  \ref{Bi_Peak}. 

For each measurement, the $^{210}$Po activity deduced from the total number of counts under the $\alpha$-peak (C$_\alpha$), as follows:
\begin{equation}
A_{Po}(t)=\frac{C_{\alpha}}{ \varepsilon_{\alpha}\; F \;t_{m}},
\end{equation}
where $t$ is the time difference between the end of the activation to the middle of the measurement time, $\varepsilon_{\alpha}$ is the NICE detector detection efficiency measured in this work  (Section \ref{NICE-DetectorCharacteristic study}), $t_m$ is the measurement time equal to 24 hours, and $F$ is a detection efficiency correction factor. This correction is needed because the Bi-sample has an extended geometry when compared to the Am-point source, and because the $\alpha$-energy is slightly lower than 5.5 MeV. The  detection efficiency correction factor for both the sample geometry and the $\alpha$-energy was calculated using Geant4-simulation and found to be $0.93\;\pm\;0.03$.

The total number of activated nuclei in each Bi-sample was determined by fitting the $^{210}$Po activity over 35 days using the function
\begin{equation}
A_{\SI{}{Po}}(t)= N_{\SI{}{activation}}\;\SI{}{\lambda_{Po}}\;[\frac{\SI{}{\lambda_{Bi}} f_{Bi}}{\SI{}{\lambda_{Po}}-\SI{}{\lambda_{Bi}}} e^{-\SI{}{\lambda_{Bi}} t}+(f_{Po}-\frac{\SI{}{\lambda_{Bi}} f_{Bi}}{\SI{}{\lambda_{Po}}-\SI{}{\lambda_{Bi}}})e^{-\SI{}{\lambda_{Po}} t}], \\
\label{po_activity}
\end{equation}
where $t$ is the time between the end of the activation and the middle of each measurement period, and $\lambda_{Bi}$ and $\lambda_{Po}$ are decay constants for $^{210}$Bi and $^{210}$Po, respectively, and $f_{Bi}$ and $f_{Po}$ as  corrections for the decaying nuclei during activation,

where 
 \begin{eqnarray}
f_{Bi}&=&\frac{1-e^{-\SI{}{\lambda_{Bi}}t_{a}}}{\SI{}{\lambda_{Bi}}t_{a}},\\
\SI{}{and}\nonumber\\
f_{Po}&=&\frac{1-e^{-\SI{}{\lambda_{Po}} t_{a}}}{\SI{}{\lambda_{Po}}t_{a}} -\frac{e^{-\SI{}{\lambda_{Bi}} t_{a}}-e^{-\SI{}{\lambda_{Po}} t_{a}}}{(\SI{}{\lambda_{Po}}-\SI{}{\lambda_{Bi}})t_{a}}.
\end{eqnarray}

Based on the fitting results, the total number of activated nuclei are
 \begin{eqnarray*}
N_{\SI{}{activation}}^{} &=&(27.62\;\pm\;0.17_{stat}\;\pm\;0.87_{sys})\;\times\;10^{7}\; \SI{} {atoms},\\   
N_{\SI{}{activation}}^{Cd} &=& (5.95\;\pm\;0.04_{stat}\;\pm\;0.19_{sys})\;\times\;10^{7}\; \SI{} {atoms}.
\end{eqnarray*}
Experimental values of the thermal cross section and resonance integral were  deduced from the measured total number of activated nuclei, and the time integrated thermal and epithermal neutron flux using the Cd-ratio difference method as follows:
 \begin{eqnarray}
\sigma_{0}&=&\frac{[\frac{^{^{210}\SI{}{Bi}}N}{^{^{209}\SI{}{Bi}}N} ]\phi_{epi}^{cd} -[\frac{^{^{210}\SI{}{Bi}}N}{^{^{209}\SI{}{Bi}}N} ]^{cd}\phi_{epi} }{\phi_{th}\phi_{epi}^{cd}-\phi_{th}^{cd}\phi_{epi}},\\
I_{0}&=&\frac{[\frac{^{^{210}\SI{}{Bi}}N}{^{^{209}\SI{}{Bi}}N} ]\phi_{th}^{cd} -[\frac{^{^{210}\SI{}{Bi}}N}{^{^{209}\SI{}{Bi}}N} ]^{cd}\phi_{th} }{\phi_{th}^{cd}\phi_{epi}-\phi_{th}\phi_{epi}^{cd}}.
 \end{eqnarray}
Using the above approach, results for the thermal cross section and resonance integral values for $^{209}$Bi(n,$\gamma$)$^{210g}$Bi were  found to be
 \begin{eqnarray*}
\sigma_{0}&=& (16.20\;\pm\;0.12_{stat}\;\pm\;0.85_{sys})\;\SI{} {mb}, \\  
 I_0&=&(89.81\;\pm\;1.07_{stat}\;\pm\;6.93_{sys})\;\SI{} {mb}.
\end{eqnarray*}

\subsection{Measurement of $\sigma_{MACS}$}\label{MACS}
 \subsubsection{Activation}
 \hfill\\
\\
A neutron beam with a quasi-stellar distribution was obtained using the $^7$Li(p,n)$^7$Be reaction.  It has been shown by Beer and Käppeler that by setting the  proton energy  to 1912 keV (30 keV above threshold), the produced neutrons will be kinematically emitted into a forward cone with a maximum opening angle of 120$^\circ$ \cite{activation}. In addition, the angle-integrated neutron energy spectrum is a good approximation of  a Maxwell$-$Boltzmann distribution, that can be used to measure the Maxwellian averaged cross section at $kT$ = 25 keV \cite{14N1}. 
 
The proton beam was obtained using the 2.0 MV Van de Graaff accelerator (VDG) at Goethe University Frankfurt, and the Li target was prepared using natural lithium material. A lithium layer with a thickness of 9.0~$\pm$~0.2~$\mu$m  was evaporated on a copper disk with  0.5 mm thickness. Two thin gold foils (Au-F and Au-B) with a disk shape that was 0.025 mm thick and 20 mm in diameter were used as flux monitors. The Bi-sample was placed  between the two gold foils, where Au-F was in front of the sample and Au-B was in back of the sample. The masses and the surface atomic density for the gold foils are given in Table \ref{tab:monitors_VDG}. 
\begin{table}[t!]
  \begin{center}
   
    \caption{Mass and the surface atomic density for the  gold foils used during  activation using the VDG accelerator.}
    \label{tab:monitors_VDG}
    \begin{tabular}{ c c c c c c c} 
    \hline
   \hline
      \multirow{2}{*}{Monitor} 	&\quad\quad	&Mass	&\quad\quad	&Diameter	&\quad\quad	&$N_{0}$   \\
                		 &\quad\quad	&     [g]  	&\quad\quad	&   [mm]  		&\quad\quad	&  [$ 10^{20}$ atoms/cm$^2$]  \\
      \hline
        && &\\
 	Au-F&\quad\quad		&  0.148 $\pm$ 0.001	&\quad\quad	&   20	&\quad\quad	&  1.44 $\pm$ 0.01\\
        && &\\
  	  Au-B&\quad\quad	&   0.157 $\pm$ 0.001	&\quad\quad	&   20	&\quad\quad	&  1.53 $\pm$ 0.01\\
   && &\\
   
    \hline
    \hline
    \end{tabular}
  \end{center}
\end{table}

\begin{figure}[!htb]
\includegraphics[scale=0.5]{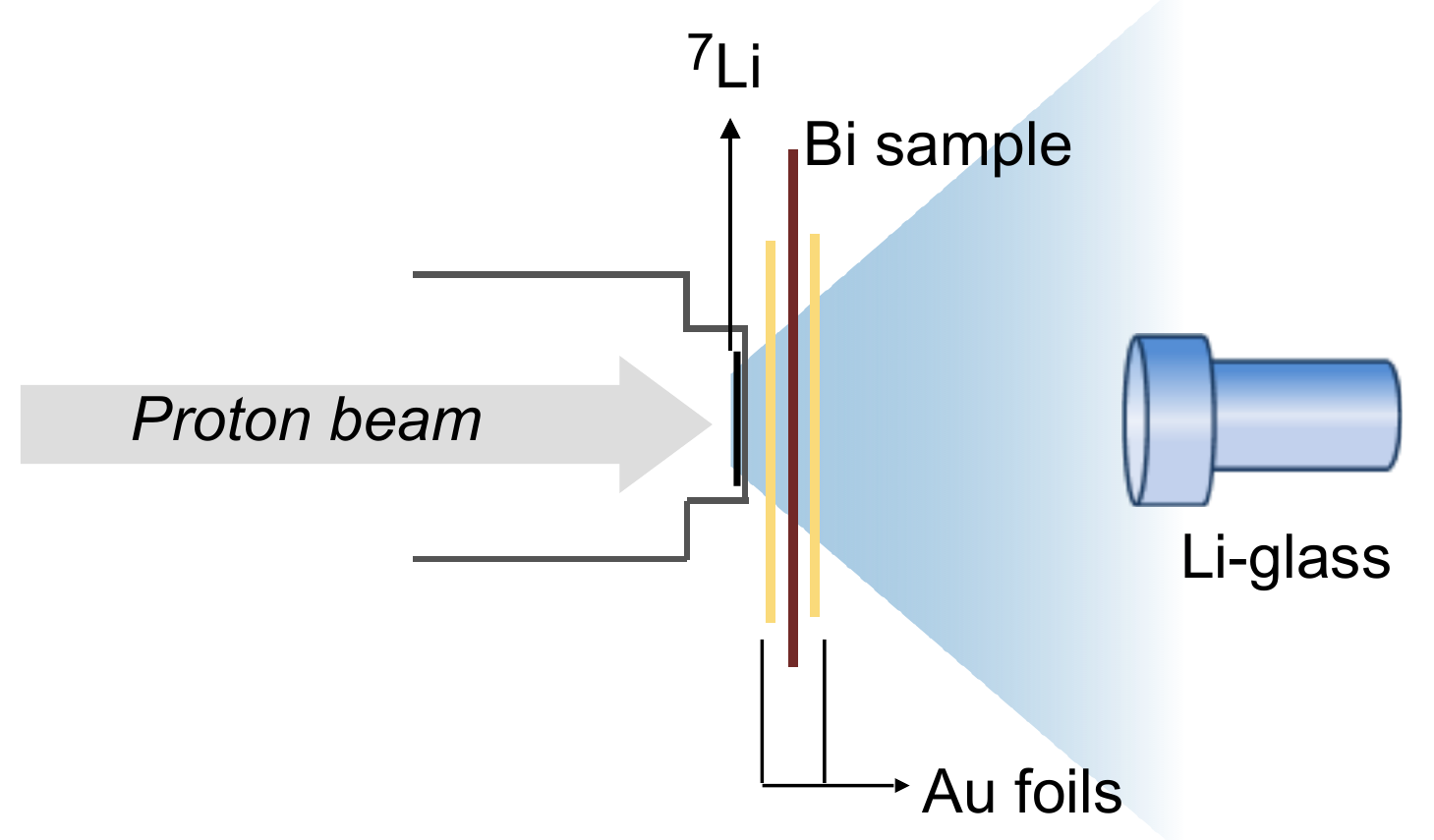}
\caption{Schematic design represent the sample, gold foils and the Li-glass detector arrangement in front of the neutron beam  (not to scale).}
\label{VDG_act}
\end{figure}

During activation,  the Bi-sample was placed at $\simeq$ 2.0~mm distance from the Li target.  At this position, the neutron cone  had a surface area of around $\simeq$ 130 mm$^2$ and will cover only the central region of both the gold and the sample foils (see Figure \ref{VDG_act}). The activation was performed for about 49 continuous hours ($t_a$ = 176787~s), and the proton beam current was kept constant at around $\simeq$13 $\mu$A. The neutron flux with time was monitored using a Li-glass detector mounted at 50 cm from the sample  at 0$^\circ$ to the beam axis. This information are important to account for the nuclei that decayed during the activation course ($f_{\SI{}{activation}}$, $f_{Bi}$, and $f_{Po}$).

At the end of the activation process, the induced activities of the gold monitors were measured using a BEGe-detector at Goethe University Frankfurt (3 in. $\times$ 3 in.). Each gold foil was fixed at 10 cm distance from the Ge-crystal and counted for around 600~s. The detection efficiency calibration process at this position was performed using a set of  calibration point-like sources ($^{57}$Co, $^{133}$Ba, $^{137}$Cs, $^{54}$Mn, and $^{60}$Co). 
The full peak detection efficiency at the measurement position was calculated to be (0.669 $\pm$ 0.005)$\%$ for the $\gamma$-line 411.8 keV. 

The total number of activated nuclei ($N_{\SI{}{activation}}$) was determined  from the number of gamma counts ($C_\gamma$) under the $\gamma$-line at 411.8 keVduring the measuring time $t_m$ as follows \cite{activation}:
\begin{equation}
N_{\SI{}{activation}}=  \frac{C_{\gamma}}{\varepsilon_{\gamma}\;I_{\gamma}\;k} \quad \frac{1}{f_{\SI{}{activation}}}  \quad\frac{1}{f_{\SI{}{waiting}}}  \quad\frac{1}{f_{\SI{}{measurement}}} ,
\label{flux_equ}
\end{equation}
where $\varepsilon_{\gamma}$ is the detector efficiency, $I_{\gamma}$ is the relative emission probability  for a particular gamma energy,  and $k$ is the dead-time correction factor. In this work,  the dead-time correction factor was determined from the ratio of  live-to-real  time, and  stayed below $2\%$. The $f_{\SI{}{activation}}, f_{\SI{}{waiting}}$, and $f_{\SI{}{measurement}}$ are corrections accounting for the fraction of  nuclei that have already decayed during the irradiation, waiting, and measurement time intervals, respectively, and they are represented as
 \begin{eqnarray}
f_{\SI{}{activation}}=\frac{\sum_i \phi_{i}\; e^{-\lambda_{Au} t_{w,i}}} {\sum_i \phi_{i}}, \nonumber\\
f_{\SI{}{waiting}}={e^{-\lambda_{\gamma} t_{w}}}, \\
f_{\SI{}{measurement}}={1- e^{-\lambda_{\gamma} t_{m}}},\nonumber 
\label{factors}
 \end{eqnarray}
where $\phi_i$ is a constant flux for each short activation interval,  and $t_{w,i}$ is the time between the end of each activation interval and the full activation time, and $\lambda_{Au}$ is the decay constant for gold, $t_{w}$ is the waiting time that represents the time interval between the end of the activation and the start of the counting process, $t_{m}$ is the measurement time, and  $\lambda_{\gamma}$ is the decay constant of the respective product nucleus. The detailed parameters and the total number of activated gold nuclei  for both monitors and the average value at the Bi sample position ($\bar N_{\SI{}{activation}}^{\SI{}{Au}}$)  are given in Table \ref{tab:flux_VDG}, which also provides the statistical and  systematic uncertainties.

\begin{table*}[t!] 
\small
  \begin{center}
    \caption{The total number of gamma counts ($C_\gamma$), and the calculated  total number of activated nuclei ($N_{\SI{}{activation}}$). }
    \label{tab:flux_VDG}
 
     \begin{ruledtabular}

    \begin{tabular}{ lcc c c c cc c} 
   
     \multirow{2}{*}{ Quantity }						&\quad\quad &\quad\quad			& \multirow{2}{*}{Au-F}					&\quad\quad &\quad\quad			& \multirow{2}{*}{Au-B}				\\
   && &&\\ 
         \hline
&& &&\\
 $C_{\gamma}$ $\pm$ $\Delta_{stat}$ [counts]	&\quad\quad &\quad\quad			&23430 $\pm$ 153 					&\quad\quad &\quad\quad		&24433 $\pm$ 156 \\
 && &&\\
  $\varepsilon$	 [$\%$]					&\quad\quad &\quad\quad	&0.669 $\pm$ 0.005					&\quad\quad &\quad\quad		&0.669 $\pm$ 0.005 \\
&& &&\\
 $f_{\SI{}{activation}}$					&\quad\quad &\quad\quad			&0.816 $\pm$ 0.001					&\quad\quad &\quad\quad		&0.816 $\pm$ 0.001 \\
&& &&\\
 $t_{w}$ [sec]							&\quad\quad &\quad\quad		&6256							&\quad\quad &\quad\quad			&6976 \\
 && &&\\
 $t_{m}$ [sec]							&\quad\quad &\quad\quad			&600 							&\quad\quad &\quad\quad		&600 \\
 && &&\\
 $k$									&\quad\quad &\quad\quad		&0.97							&\quad\quad &\quad\quad			&0.97 \\
&& &&\\
 $N_{\SI{}{activation}}^{\SI{}{Au}}\pm\; \Delta_{stat}\pm\;\Delta_{sys}$ [$10^9$ atoms]&\quad\quad &\quad\quad			&2.64 $\pm$ 0.02 $\pm$ 0.03 	&\quad\quad &\quad\quad 		&2.76 $\pm$ 0.02 $\pm$ 0.04 \\
&& &&\\
&& &&\\

 $\bar N_{\SI{}{activation}}^{\SI{}{Au}}\pm\; \Delta_{stat}\pm\;\Delta_{sys}$ [$10^9$ atoms]&\quad\quad 	&\quad\quad		&2.70 $\pm$ 0.03 $\pm$ 0.05 	&\quad\quad &\quad\quad 		& \\
&& &&\\
  
    \end{tabular}
      \end{ruledtabular}

 \end{center}
\end{table*}

\begin{figure}[!htb]
\centering 
\includegraphics[scale=0.5]{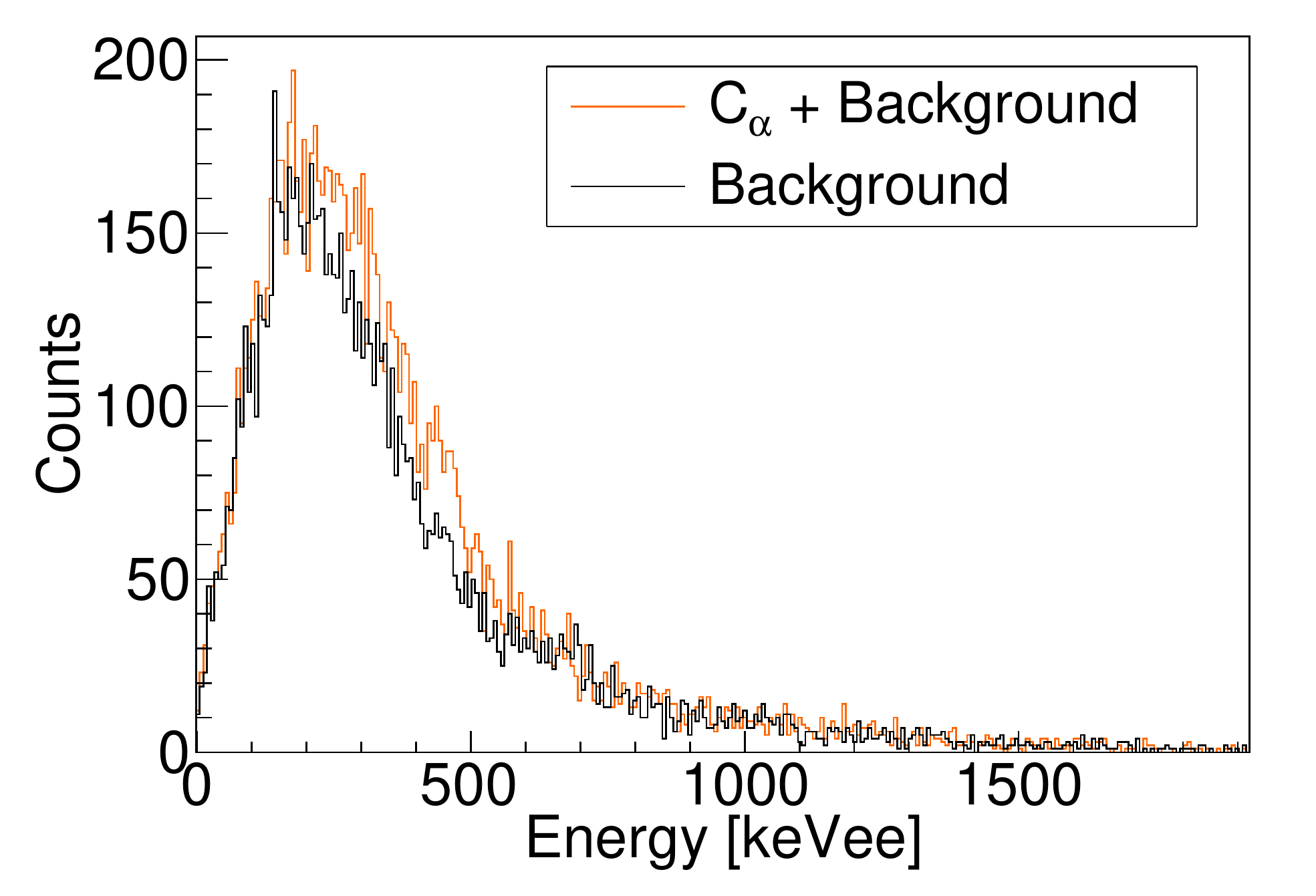}
\caption[The peak obtained using the NICE detector ($C_\alpha$ and background) for the activated Bi-sample in comparison to the ambient background spectrum. The measurement started 55 days after the end of the activation for one day.]{The peak obtained using the NICE-detector ($C_\alpha$ and background) for the activated Bi-sample in comparison to the ambient background spectrum. The measurement started 55 days after the end of the activation for one day.}
\label{fig:Bi_Bg_VDG}
\end{figure}

The Bi-sample was counted using the NICE detector. The counting process started 55 days after the end of the activation. In this activation the count rate was low, as it was limited by both the neutron flux and the activation time. An example of the peak obtained using the NICE detector ($C_\alpha$ and background) compared to the ambient background spectrum which was measured independently is shown in Figure \ref{fig:Bi_Bg_VDG}. The measurement started 55 days after the end of the activation and for one day. The low count rate combined with the poor resolution of the NICE detector made it difficult to distinguish the $\alpha$-peak from the relatively high background count rate. At this stage, the total number of activated nuclei was determined by fitting the total number of counts under the peak using the following function

\begin{widetext}
\begin{equation}
\Delta N(t)= N_{\SI{}{activation}}\;\varepsilon_{\alpha} \;F\; t_m\;\SI{}{\lambda_{Po}}\;[\frac{\SI{}{\lambda_{Bi}} f_{Bi}}{\SI{}{\lambda_{Po}}-\SI{}{\lambda_{Bi}}} e^{-\SI{}{\lambda_{Bi}} t}+(f_{Po}-\frac{\SI{}{\lambda_{Bi}} f_{Bi}}{\SI{}{\lambda_{Po}}-\SI{}{\lambda_{Bi}}})e^{-\SI{}{\lambda_{Po}} t}] + \SI{}{C}, \\
\label{po_activity}
\end{equation}
\end{widetext}
where $t$ is the time at the middle of each measurement, $\Delta N (t)$ is the total number of counts under the peak  ($C_\alpha$ and background), $\varepsilon_{\alpha}$ is the NICE detector efficiency, and $F$ efficiency correction factor, $t_m$ is the measurement time, $\SI{}{\lambda_{Bi}}$ and $\SI{}{\lambda_{Po}}$ are the decay constant for $^{210g}$Bi and $^{210}$Po, respectively, and $f_{Bi}$ and $f_{Po}$ are corrections that account  for the $^{210g}$Bi and $^{210}$Po nuclei that decayed during the activation course, respectively, and C is a constant that accounts for the contribution coming from the ambient background. 
Similarly, $f_{Bi}$ and $f_{Po}$ were determined by considering the full activation interval ($t_a$) as a sequence of consecutive short activation intervals ($t_i$), each with a constant neutron flux ($\phi_i$). Accordingly, the correction factors can be expressed as 
 \begin{eqnarray}
f_{Bi}&=&\frac{\sum_i \;\phi_i \;e^{-\lambda_{Bi} t_{w,i} }}{ \sum_i \;\phi_i },\\
f_{Po}&=&\frac{\lambda_{Bi}} {\lambda_{Po}-\lambda_{Bi}} \frac{\sum_i \phi_i\; e^ {-\lambda_{Po} t_a} \;[e^{(\lambda_{Po}-\lambda_{Bi}) t_a}-e^{(\lambda_{Po}-\lambda_{Bi}) t_i}] }{\sum_i \phi_i\; } .
 \end{eqnarray}
Where $t_a$ is the total activation time, and  $t_i$ is the time at the end of each activation interval, and $t_{w,i}$ is the time between the end of each activation interval and the full activation time.

\begin{figure}[]
\centering 

\includegraphics[scale=0.5]{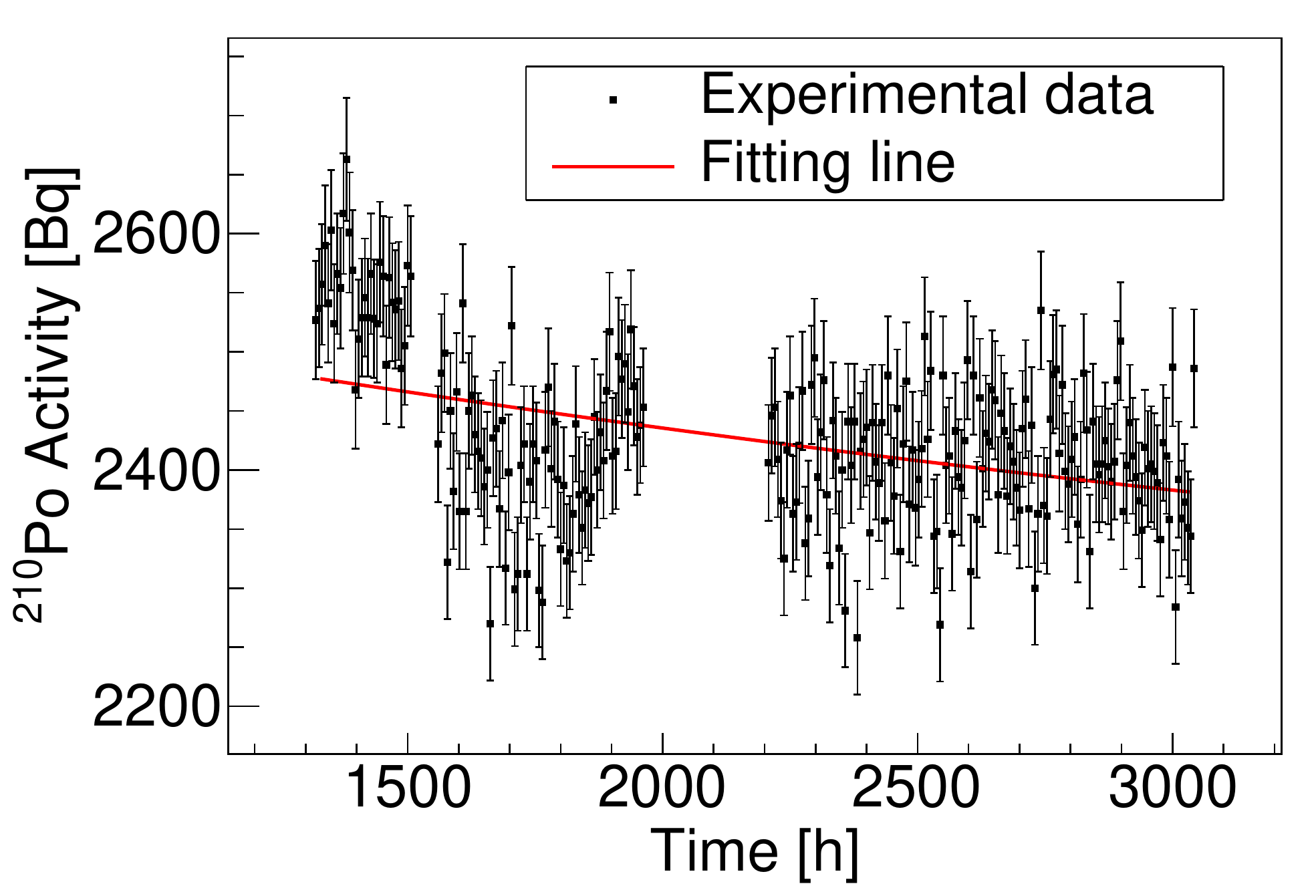}
\caption[The total number of counts behaviour over time for the activation using VDG accelerator.]{The total number of counts behaviour over time. Each point presents one measurement of 6 hours and is placed at the middle of each measurement. The error bars represent the statistical uncertainty of each measurement.}
\label{fig:activity_fit_VDG}
\end{figure}

In this work,  240 measurements each for 6 hours were performed between day 55 and day 126 after the end of the activation. Figure \ref{fig:activity_fit_VDG} shows the experimental data of the total number of counts and the fitting curve. Based on the fitting results, the total number of activated nuclei and the estimated  background  are  
 \begin{eqnarray*}
N_{\SI{}{activation}}^{\SI{}{Bi}} &=& ( 7.65\;\pm\;0.82_{stat}\;\pm\;0.45_{sys} )\;\times\;10^{5}\; \SI{} {atoms},\\   
\SI{}{C} &=&( 2156\;\pm\; 29)\; \SI{}{counts} .\\
 \end{eqnarray*}

One can deduce the spectrum averaged cross sections of $^{209}$Bi(n,$\gamma$)$^{210g}$Bi,  for the corresponding quasi$-$Maxwellian spectrum at $kT$ = 25 keV  ($\sigma_{SACS,\;Bi}$), by comparing the total number of activated nuclei from both the bismuth sample and the gold monitor, each normalized to its initial number of atoms, and  multiplied with the spectrum averaged cross sections of gold as follows:
 \begin {equation}
\sigma_{\SI{}{SACS,Bi}}= \sigma_{\SI{}{SACS,Au}}\;\;\frac{N_{\SI{}{activation}}^{\SI{}{Bi}}}{{N_0^{\SI{}{Bi}}}}\;\;\frac{N_{0}^{\SI{}{Au}}}{{N_{\SI{}{activation}}^{\SI{}{Au}}}},
\label{compareeq}
\end{equation}
where $N_{\SI{}{activation}}^{\SI{}{Bi}}$ and $N_{\SI{}{activation}}^{\SI{}{Au}}$ are the total number of activated nuclei for Bi and Au samples, respectively, and ${N_0^{\SI{}{Bi}}}$ and ${N_0^{\SI{}{Au}}}$ are the surface atomic density (atoms/cm$^{2}$) for Bi and Au samples,  respectively, and $\sigma_{\SI{}{SACS,Au}}$ is the spectrum averaged cross sections for $^{197}$Au(n,$\gamma$)$^{198}$Au for the  corresponding experimental quasi$-$Maxwellian spectrum at $kT$ = 25 keV. 
The experimental cross section for $^{197}$Au(n,$\gamma$)$^{198}$Au  was obtained using PINO simulation code \cite{PINO}. The simulation was performed twice for each gold foil and the average value was found to be  0.651~$\pm$~0.006~b.

Using the above approach (Equation \ref{compareeq}), the  spectrum averaged cross section of the $^{209}$Bi(n,$\gamma$)$^{210g}$Bi reaction was calculated and found to be 
 \begin {equation*}
\sigma_{\SI{}{SACS,Bi}} = (1.83\;\pm\;0.20_{stat}\;\pm\;0.15_{sys})\;\SI{}{mb}.
\end{equation*}
The relatively high statistical uncertainty ($\simeq\;$10\;$\%$) is due to the low count rate obtained during the measurement. While the $\simeq\;$8\;$\%$ systematic uncertainty accounts for the uncertainty  in all the initial parameter (e.g.  initial masses, $t_{1/2}$, $t_w$ and $t_{m}$). In addition, a 5$\%$ systematic uncertainty was assumed to account for the constant background assumption during the fitting process. 

\subsubsection{Calculations of the Maxwellian Averaged cross section ($\sigma_\mathrm{MACS}$)}
\hfill\\
\\
However,  the experimental spectrum obtained using the $^7$Li(p,n)$^7$Be reaction corresponds in  good approximation to a Maxwell$-$Boltzmann  spectrum for thermal energy of $kT$~=~25~keV. But, the cutoff energy at 110 keV in the experimental spectrum implies that the contributions from higher neutron energies are not included. Therefore, in order to deduce an accurate value of  the Maxwellian average cross section for $^{209}$Bi(n,$\gamma$)$^{210g}$Bi denoted as  $\sigma_\mathrm{MACS,\;Bi}$, from the  measured spectrum averaged cross sections $\sigma_\mathrm{{SACS,Bi}}$. Or, if the $\sigma_\mathrm{MACS,\;Bi}$ is needed to be  extrapolated for different temperatures ($kT$), a final correction to the measured cross section value is required: 
\begin {equation}
\sigma_\mathrm{MACS, Bi}(kT)= \sigma_\mathrm{SACS, Bi} \; C_f (kT).
\end{equation}
where $C_f (kT)$ is the correction factor. To determine the correction factor, one can use the available differential cross section data ($\sigma_{n,\gamma} (E_n)$) from the evaluated data libraries and employ the PINO simulation code. In this work the ENDF/B-VIII.0  data libraries was used  \cite{ENDF1}, and was folded twice, once with the experimental neutron  distribution using PINO simulation code, and once with a typical Maxwell$-$Boltzmann distribution, to  calculate the  $\sigma_\mathrm{SACS}^\mathrm{ENDF}$ and $\sigma_\mathrm{MACS}^\mathrm{ENDF}(kT)$ , respectively. Thereby, the correction factor is the ratio of  $_\mathrm{MACS}^\mathrm{ENDF}(kT)$ to $\sigma_\mathrm{SACS}^\mathrm{ENDF}$ expressed mathematically as follows:

 \begin{eqnarray}
C_f (kT)&=& \frac{\sigma_\mathrm{MACS}^\mathrm{ENDF}(kT)}{\sigma_\mathrm{SACS}^\mathrm{ENDF}}, \nonumber \\
\sigma_\mathrm{MACS}^\mathrm{ENDF}(kT)&=&\frac{2}{\sqrt{\pi}}\frac{1}{(kT)^2} \int_0^\infty E_n\sigma^\mathrm{ENDF}(E_n){e}^{(\frac{-E_n}{kT})}\mathrm{d}E_n, \\
\sigma_\mathrm{SACS}^\mathrm{ENDF}&=& \frac{\int_0^\infty \;\sigma^\mathrm{ENDF}(E_n)\;\phi_\mathrm{exp}(E_n)\;\mathrm{d}E_n}{ \int_0^\infty \phi_\mathrm{exp}(E_n)\;\mathrm{d}E_n}.\nonumber
\\ \nonumber
\label{CF}
 \end{eqnarray}

Where $E_n$ is the neutron energy in the centre-of-mass system, and $\sigma^\mathrm{ENDF}$ is the differential capture cross section from the ENDF data libraries, and $\phi_{exp}(E_n)\;dE_n$ is the experimental neutron distribution between $E$ and $E+dE$. 

In this work, the $\sigma_\mathrm{MACS}^\mathrm{ENDF}$  was calculated using the ROOT-toolkit \cite{ROOT}, the deferential cross section was folded from 0~eV to 10~MeV with energy step size 1~eV, and extrapolated  to different thermal temperatures from 5 keV to 100 keV. While the $\sigma_\mathrm{SACS}^\mathrm{ENDF}$ was measured using the PINO simulation code and found to be 3.07~mb. 

To follow the above approach for calculating the correction factor and the corresponding  $\sigma_\mathrm{MACS, Bi}$, one should keep in mind that the cross section values obtained from the ENDF data library are the total capture cross section, which feed both the ground state ($^{210g}$Bi) and the long lived isomeric state ($^{210m}$Bi). Thereby, the folding technique with the Maxwell-Boltzmann distribution and with the experimental distribution, will provide the total capture cross section ($\sigma_{tot} = \sigma_{g}+\sigma_{m}$). While in the experiment, and by measuring the $^{210}$Po activity one can only determine the partial capture cross section to the ground state ($\sigma_{g}$). 

Therefore, an accurate knowledge of the  the energy dependence of the isomeric ratio  ($\sigma_g$/$\sigma_m$) is essential. The experimental data available for this ratio are limited,  Borella \textit{et al.} calculated a value of 1.17 $\pm$ 0.05 with thermal neutrons \cite{Borella}, and Saito \textit{et al.} reported a value of 2.98 $\pm$ 1.92  and 0.81 $\pm$ 0.25 at  30 keV and 534 keV, respectively \cite{Saito1}.  The ENDF data library does not give any information about the isomeric ratio, but other data libraries such as JEFF-3.2 and RUSFOND-2010  provide a constant ratio up to 1 MeV  neutron energy \cite{isomericratio}. 

In this work, the correction factor was calculated with the assumption that the isomeric ratio $\sigma_g/\sigma_m$ is constant.  The values of the calculated $\sigma_\mathrm{MACS}^\mathrm{ENDF}$, the corresponding correction factor ($C_f (kT)$), and the calculated experimental $\sigma_\mathrm{MACS, Bi}$ are listed in Table \ref{tab:lMACS}.
\begin{table}[t!]

  \begin{center}
 \caption[The total Maxwellian averaged capture cross section calculated using the ENDF data libraries. And the Maxwellian average cross section for $^{209}$Bi(n, $\gamma$)$^{210g}$Bi calculated in this work.]{ The total capture cross-section calculated using the ENDF data libraries. And the Maxwellian average cross section for $^{209}$Bi(n,$\gamma$)$^{210g}$Bi reaction calculated in this work. The systematic and statistical uncertainties  are equal to those obtained from the measurement  at $kT$ = 25 keV.}
   \label{tab:lMACS}

    \begin{tabular}{cccc ccccccc} 
    \hline
        \hline
${kT}$&\quad\quad		&{$\sigma_\mathrm{MACS}^\mathrm{ENDF}$}				&		&\multirow{2}{*}{${C_F}$}	&\quad\quad\quad&  {{$\sigma_\mathrm{MACS,Bi} \pm \Delta_{stat}  \pm \Delta_{sys}$}} \\
{[keV]}&\quad\quad &             {[mb]}                     & &      		             &\quad\quad&{[mb]}  \\

\hline
&&&\\
{5}                 &\quad     &{15.25}      &              &{4.96}&\quad&          {9.08 $\pm$ 0.99 $\pm$ 0.73}\\

{10}                 &\quad     &{7.76}      &                &{2.52}&\quad&          {4.62 $\pm$ 0.50 $\pm$ 0.37}\\
{15}                 &\quad     &{5.21}      &                &{1.70}&\quad&          {3.10 $\pm$ 0.34 $\pm$ 0.25}\\
{20}                 &\quad     &{4.07}      &                &{1.33}&\quad&          {2.43 $\pm$ 0.26 $\pm$ 0.19}\\
{25}                 &\quad    &{3.57}     &                &{1.16}&\quad&          {2.12 $\pm$ 0.23 $\pm$ 0.17}\\

{30}                 &\quad     &{3.37}      &                &{1.09}&\quad&          {2.01 $\pm$ 0.22 $\pm$ 0.16}\\

{35}                 &\quad    &{3.33}      &                &{1.08}&\quad&          {1.98 $\pm$ 0.22 $\pm$  0.16}\\

{40}                 &\quad    &{3.36}      &                &{1.09}&\quad&          {2.00 $\pm$ 0.22 $\pm$  0.16}\\

{45}                 &\quad    &{3.42}     &                &{1.11}&\quad&          {2.04 $\pm$ 0.22 $\pm$  0.16}\\

{50}                 &\quad    &{3.50}      &                &{1.14}&\quad&          {2.08 $\pm$ 0.23 $\pm$  0.17}\\
{55}                 &\quad     &{3.56}      &                &{1.16}&\quad&          {2.12 $\pm$ 0.23 $\pm$ 0.17}\\

{60}                 &\quad    &{3.62}      &                &{1.18}&\quad&          {2.16 $\pm$ 0.24 $\pm$ 0.17}\\
{65}                 &\quad     &{3.68}      &                &{1.20}&\quad&          {2.19 $\pm$ 0.24 $\pm$ 0.18}\\
{70}                 &\quad     &{3.72}      &                &{1.21}&\quad&          {2.21 $\pm$ 0.24 $\pm$ 0.18}\\
{75}                 &\quad     &{3.75}      &                &{1.22}&\quad&          {2.24 $\pm$ 0.24 $\pm$ 0.18}\\
{80}                 &\quad   &{3.78}      &                &{1.23}&\quad&          {2.25 $\pm$ 0.25 $\pm$ 0.18}\\
{85}                 &\quad    &{3.80}      &                &{1.24}&\quad&          {2.25 $\pm$ 0.25 $\pm$ 0.18}\\

{90}                 &\quad     &{3.82}      &                &{1.24}&\quad&          {2.27 $\pm$ 0.25 $\pm$ 0.18}\\
{95}                 &\quad    &{3.83}      &                &{1.24}&\quad&          {2.28 $\pm$ 0.25 $\pm$ 0.18}\\
{100}                 &\quad   &{3.83}      &                &{1.25}&\quad&          {2.28 $\pm$ 0.25 $\pm$ 0.18}\\
&&&\\
 \hline
\hline
\end{tabular}

  \end{center}
\end{table}

\section{Summary}
The neutron capture cross section of $^{209}$Bi(n,$\gamma$)$^{210g}$Bi was calculated using the new detector setup (NICE). This detector is build and designed to be used in experiments of neutron induced reaction with a charged particle in the exit channel. First part of this work was to explore the performance of the NICE detector using calibration sources. This investigation showed that the NICE detector setup  and the applied  time coincidence technique are capable to measure $\alpha$-particles with sufficient efficiency. The second part was to measure  the $^{209}$Bi(n,$\gamma$)$^{210g}$Bi  cross section 
at three different energies, including thermal capture cross section($\sigma_0$), resonance integral ($I_0$) and the Maxwellian average cross section at stellar energy of $kT$~=~30 keV ($\sigma_{MACS}$). 
\begin{table*}[!htb]
  \begin{center}
 \caption[An overview of the thermal cross sections of $^{209}$Bi(n,$\gamma$) $^{210}$Bi reaction.]{An overview of the thermal cross sections of $^{209}$Bi(n,$\gamma$)$^{210}$Bi reaction. Evaluated data are reported as total cross section ($g$+m), while experimental data is measured for the ground state ($g$). The uncertainties are given in brackets. }
   \label{tab:listdata1}
        \begin{ruledtabular}

    \begin{tabular}{lccc ccccccc} 

\multirow{2}{*}{ {Reference}}&\quad\quad&\multirow{2}{*}{ {Method, Detection}}&\quad\quad&$ {\sigma_{g+m}}$&\quad\quad&$ {\sigma_{g}}$&\quad\quad& $ {\SI{}{I}_{g+m}}$&\quad\quad& \multirow{2}{*}{ {Ref.}} \\
                                        &\quad\quad &                                  &\quad\quad& {[mb]}       		  &\quad\quad&  {[mb]}             &\quad\quad& { [mb]}         &\quad\quad& \\

\hline
&&&&&&\\

 {Seren \textit{et al.} (1947)}                 &\quad    & {Activation, $\beta$}      &\quad\quad&                 &\quad\quad& {15.0 (2.0)}&\quad\quad&              &\quad\quad&\cite{seren}\\

 {Colmer and Littler (1950)}        &\quad     & {Activation, $\alpha$}     &\quad\quad&                 &\quad\quad& { 20.5 (1.5)}&\quad\quad&        &\quad\quad&\cite{Colmer}\\

 {Takiue and Ishikawa (1978)}   &\quad   & {Activation, $\beta$}       &\quad\quad&                  &\quad\quad& {24.2 (0.4)}&\quad\quad&         &\quad\quad&\cite{Takiue}\\

 {Letourneau \textit{et al.} (2006)}        &\quad    & {Activation, $\alpha$}      &\quad\quad&                 &\quad\quad& {16.08 (1.8)}&\quad\quad&     &\quad\quad &\cite{Letourneau}\\ 

 {Letourneau \textit{et al.} (2006)}       &\quad    & {Activation, $\gamma$}    &\quad\quad &                 &\quad\quad& {18.04 (0.9)}      & \quad\quad   &                & \quad\quad& \cite{Letourneau}\\
&&&&&&\\
 {ENDF/B-VII.1}          &\quad    & {Evaluation}&\quad\quad       & {33.8}&\quad\quad&           &\quad\quad   & {205.0}&\quad\quad    &\cite{ENDF} \\
 {JENDL-4.0 (2011)}   &\quad    & {Evaluation}&\quad\quad       & {34.2} &\quad\quad&          & \quad\quad  & {171.9}& \quad\quad   &\cite{JENDL}\\ 
&&&&&&\\
 {This work (NICE-detector)}      &\quad    & {Activation, $\alpha$}     &\quad\quad&                 &\quad\quad& { 16.20 (0.97)}&\quad\quad&         &\quad\quad&\\

    \end{tabular}
          \end{ruledtabular}

  \end{center}
\end{table*}

According to our knowledge, this is the first experimental measurement of the resonance integral, while several measurements of the thermal cross section have been reported in previous studies. 
Table \ref{tab:listdata1} lists a number of evaluated and  experimental values obtained using the activation technique,~and~compared~to~the~cross ~sections~obtained~in~this~work.

Based on measuring the $^{210}$Bi induced activity and by counting $\beta$-particles, Seren \textit{et al.} \cite{seren} and Takiue and Ishikawae \cite{Takiue} reported two different values. Colmer and Littler \cite{Colmer} and Letourneau \textit{et al.} \cite{Letourneau} obtained the cross section values by measuring $^{210}$Po induced activity, and the low intensity gamma-ray coming from the de-excitation of $^{206}$Pb (Letourneau \textit{et al.}). 
 
Within the obtained  uncertainty, the measured cross section  in this work is in  good agreement with Seren \textit{et al.} and Letourneau \textit{et al.},  but 20$\%$ lower than Colmer and Littler and  30$\%$  than Takiue and Ishikawa. 
%

\begin{table*}[!htb]
  \begin{center}
 \caption[An overview of the Maxwellian avargae cross sections at $kT$ = 30 keV of $^{209}$Bi(n, $\gamma$)$^{210g}$Bi reaction.]{An overview of the experimental data of cross sections at $kT$ = 30 keV of $^{209}$Bi(n,$\gamma$)$^{210g}$Bi reaction. The uncertainties are given in brackets. }
   \label{tab:listdata2}
      \begin{ruledtabular}

    \begin{tabular}{lccc cccc} 
   \multirow{2}{*}{ {Reference}}&\quad & \multirow{2}{*}{ {Method, Detection}}&\quad&&$ {\sigma_{g}}$&\quad&  \multirow{2}{*}{ {Ref.}} \\
                                        &\quad&                                  &\quad& &{[mb]}       		             &\quad& \\

\hline
&&&&&&\\


 {Ratzel \textit{et al.} (2004)}                 &\quad   & {Activation, $\beta$}      &\quad&                & {2.54  (0.14)}&\quad&           \cite{Ratzel}\\

 {Bisterzo \textit{et al.} (2008)}        &\quad     & {Activation, $\alpha$}     &\quad&                \quad& {2.16 (0.07) }&\quad&     \cite{Bisterzo}\\

 {Shor \textit{et al.} (2017) }        &\quad    & {Activation, $\alpha, \beta, \gamma$}      &\quad&                 \quad& {1.84 (0.09)}&\quad&   \cite{Shor}\\ 

&&&&&&\\
 {This work (NICE-detector)}      &\quad    & {Activation, $\alpha$}                                &\quad&                \quad& {2.01\;(0.38)}&\quad&         \quad \\

\end{tabular}
   \end{ruledtabular}

        \end{center}
\end{table*}
A comparison between the Maxwellian average cross section at a thermal energy of 30~keV  measured in this work and the values  that have been reported in previous studies is given in Table \ref{tab:listdata2}. Based on the activation technique, Ratzel \textit{et al.}   and Bisterzo \textit{et al.}  obtained  the cross section value by measuring the  $^{210}$Bi and the $^{210}$Po  induced activities, respectively  \cite{Ratzel,Bisterzo}. In 2017 Shor \textit{et al.} calculated the cross section by measuring  the $^{210g}$Bi, the $^{210}$Po  induced activities, and the low intensity gamma ray coming from the de-excitation of $^{206}$Pb \cite{Shor} and reported value of 1.84 $\pm$ 0.09 mb . 

Within the uncertainty obtained in this work, the calculated cross section is in agreement with the reported ones. The discrepancies between the previous works could not be resolved with this experiment, but the potential of the new detection technique could be shown. In a future work, the uncertainties in this cross section can easily be reduced, since they originate largely from statistical uncertainties and sample properties. Higher statistics can be achieved by  increasing the neutron flux or the total activation time. 

\section*{Acknowledgment}
 Authors gratefully acknowledge the financial support by the DFG-project NICE (RE 3461/3-1) and HIC for FAIR. A special thanks are due to the staff of the TRIGA-reactor, Mainz, and the staff of the VDG accelerator at Goethe University Frankfurt. 
\section*{References}


\end{document}